\documentclass[useAMS,usenatbib]{mn2e}

\usepackage{lscape}
\usepackage{longtable}
\usepackage{amssymb}

\usepackage{graphicx}

\title{Optical Galaxy Clusters in the Deep Lens Survey}
\author[B. Ascaso, D. Wittman, W. Dawson]{B. Ascaso$^{1,2}$\thanks{E-mail:
ascaso@iaa.es};  D. Wittman$^{2}$ ;  W. Dawson$^{2}$\\ 
$^{1}$Instituto Astrofisica Andaluc\'ia (CSIC). Glorieta Astronomia s/n, 18008, Granada, Spain\\
$^{2}$Physics Dpt. One Shields Av. Davis, CA 95616, USA}
\begin{document}

\date{Accepted . Received }


\maketitle

\label{firstpage}

\begin{abstract}
We present the first sample of 882 optically selected galaxy clusters in the Deep Lens Survey (DLS), selected with the Bayesian Cluster Finder.  We create mock DLS data to assess completeness and purity rates, and find that both are  at least 70\% within 0.1$\le z \le$ 1.2 for clusters with $M_{200}\ge 1.2\times 10^{14}M_{\odot}$.  We verified the integrity of the sample by performing several comparisons with other optical, weak lensing, X-ray and spectroscopic surveys which overlap the DLS footprint: the estimated redshifts are consistent with the spectroscopic redshifts of known clusters (for $z>0.25$ where saturation in the DLS is not an issue); our richness estimates in combination with a previously calibrated richness-mass relation yields individual cluster mass estimates consistent with available SHeLS dynamical mass estimates; synthetic mass maps made from the optical mass estimates are correlated ($>3\sigma$ significance) with the weak lensing mass maps; and the mass function thus derived is consistent with theoretical predictions for the CDM scenario.  With the verified sample we investigated correlations between the brightest cluster galaxies (BCG) properties and the host cluster properties within a broader range in redshift (0.25 $\le z \le$ 0.8) and mass ($\ge2.4\times 10^{14}M_{\odot}$) than in previous work. We find that the slope of the BCG magnitude-redshift relation throughout this redshift range is consistent with that found at lower redshifts.  This result supports an extrapolation to higher  redshift of passive evolution of the BCG within the hierarchical  scenario. 
\end{abstract}


\section{Introduction}

Galaxy clusters are important probes of cosmology and galaxy evolution because they are the largest virialized structures in the universe, they occupy very massive dark matter halos, and they provide a unique physical environment for the transformation of galaxies.  Because their abundance is extremely sensitive to several cosmological parameters \citep{robertson09,stanek09} they can serve as effective cosmological probes. While many works have provided cosmological constraints with massive-low redshift samples \citep{henry00,mantz08,mantz10}, present and future surveys are obtaining samples spanning a wide range of redshift and mass (e.g. \citealt{vikhlinin09,rozo10,allen11}) in order to capture evolution and provide constraints on dark energy evolution and mass-richness relation. 

Galaxy clusters, being the densest environments, also act as laboratories for a number of studies related to galaxy evolution. For instance, many studies have found an evolution of the blue fraction with redshift albeit with a wide dispersion \citep{butcher84,margoniner01,depropris04,ascaso08}, measured and fit the luminosity function down to the faint end for different surveys \citep{blanton03,harsono09,depropris13}, characterized the slope of the color-magnitude relation to at least up to redshift 1.6 \citep{mei06,ascaso08,mei09,papovich10}, or measured the  dependence of the star formation rate with the environment and  redshift \citep{lewis02,balogh04,ma08}.

Large-area surveys have been able to detect very rare and massive clusters (M$>2-3\times10^{15}M_{\odot}$) with a variety of different methods (e.g., SDSS \citealt{york00}, MACS, \citealt{ebeling01,ebeling10}; Planck, \citealt{planck11a}). In general, the most massive clusters are easy to identify up to moderate redshift (z$\lesssim$ 0.5) since they contain large numbers of tightly clustered galaxies  \citep{abell89,depropris02,ascaso08}, strong X-ray emission signatures  \citep{ebeling96,ebeling01,rosati02,bohringer04,bohringer07}, relatively strong features in the gravitational lensing shear field  \citep{wittman06,limousin07,postman12a} and potential Sunyaev-Zel'dovich (SZ) signatures \citep{ascaso07,diego08,menanteau09,menanteau10}. Moreover, many high redshift clusters have also been detected in different surveys with a number of different techniques  (e.g., optical-infrared: RCS2319530038.0 at z=0.9, \citealt{gladders05,gilbank08}; RzCS052 at z=1.016, \citealt{andreon08}; ISCS J143809+341419 at z=1.41, \citealt{stanford05}; a structure at z$\sim$ 1.6, \citealt{trevese07,castellano07}; JKCS 041 at z=1.803, \citealt{andreon09}; X-rays: ClJ1226.9+3332 at z=0.89. \citealt{ebeling01}; MACSJ0744.9+3927 at z=0.69, \citealt{ebeling07}; XMMU J2235.3-2557 at z=1.39, \citealt{mullis05}; SZ effect: ACT-CLJ0102-4915 'El Gordo' at z=0.87, \citealt{menanteau12}; PLCK G266.6-27.3 at z$\sim 1$, \citealt{planck11b}). All these techniques depend strongly on the depth of the observations and the characterization of their selection functions for each particular dataset is crucial to compare and avoid systematic errors.

In recent years, the low redshift regime of optical clusters has been widely sampled in the literature. Very wide surveys with spectroscopy such as the Sloan Digital Sky Survey (SDSS,\citealt{york00}) have been explored resulting in several cluster catalogs \citep{koester07,szabo11,hao10,wen12}. These surveys probe up to z$\sim$ 0.45 (see Figure 3 in \citealt{wen12}) and down to masses of a few $10^{13}M_{\odot}$.  As for high redshift cluster samples, two main surveys with deep infrared data have been widely explored with a systematic search of galaxy clusters: the Spitzer Infrared Array Camera Shallow Survey \citep{eisenhardt08} and the Spitzer Adaptation of the Red-Sequence Cluster Survey (SpARCS; \citealt{wilson06}). These surveys have discovered several clusters at z$>$1, some of them confirmed spectroscopically.

Regarding optical cluster catalogs within the intermediate redshift range, $0.3<z<1.0$, a number of surveys have been released during the last years.  Their width, depth and photometric redshift accuracy play an important role in the resulting number of clusters and the limits in mass and redshift. Extremely deep surveys such as GOODS \citep{giavalisco04} or COSMOS  \citep{scoville07} also have small sampling areas, achieving a similar cosmic volume to shallower and wider surveys such as the Canada-France-Hawaii-Telescope Legacy Survey  (CFHTLS). In a previous paper, (\citealt{ascaso12}, A12 hereafter), we detected galaxy clusters in the CFHTLS-Archive-Research Survey (CARS, \citealt{erben09}), finding agreement with other works previously made in the CFHTLS-Wide and Deep.  The resulting sample extended up to z$<$1.2 and masses down to $4\times10^{14}M_{\odot}$. Very few surveys at this range of redshift are both deep and wide, which makes difficult the identification of large samples of clusters down to small masses at moderate redshifts.  

The Deep Lens Survey (DLS; \citealt{wittman02}), a very deep four optical band survey of 20 square degrees, is both wide and deep enough to obtain a substantial number of clusters down to $\sim 2\times 10^{14} M_{\odot}$ in this redshift range.  In A12 we introduced a new method to detect galaxy clusters, the Bayesian Cluster Finder (BCF), based on a Bayesian approach of the matched filter technique. We tested the algorithm extensively on simulations, finding higher rates of both completeness and purity than other methods. 

In this paper, we apply the BCF to the DLS and release the first optical galaxy cluster catalog in the survey including positions, redshift, richnesses, masses and radii. The number of clusters per volume unit considered here is significantly higher than SDSS, especially at z$>$0.2. Furthermore, we are obtaining a similar density of clusters as similar-depth surveys (CFTHLS-Deep, for instance) but with a considerable increase (five times) in area.  Sample variance is nevertheless substantial in a 20 $deg^2$ survey, so we do not attempt to constrain cosmology here.  Our primary goal is to select a cluster sample to enable a series of studies of the galactic population in such clusters. In this first work, we investigate the relation between the BCG properties of the sample with the properties of the host clusters within a redshift range of  0.25$\le z \le$0.8 and masses M$>2.4 \times 10^{14} M_{\odot}$. This analysis is the first one carried out with such a broad sample and will extend the previous results for the relationships between BCG and host cluster  \citep{ree07,ascaso11,wen12}.

The structure of the paper is as follows. In Section 2, we describe the DLS data we are using in this paper. Section 3 is devoted to the DLS optical cluster detections and the comparison with other detections in the same survey. In Section 4, we analyze the mass and redshift properties of the detections and in Section 5, we study the relationship between the BCGs and the host cluster properties. Finally, Section 6 includes the summary with the final conclusions of the paper. Where appropriate, we use $H_0$=71 km s$^{-1}$ Mpc$^{-1}$, $\Omega_M$ =0.27, $\Omega_L$=0.73 throughout this paper.

\section{The Deep Lens Survey}

The Deep Lens Survey (DLS, \citealt{wittman02})  is a very deep {\it BVRz} imaging survey of five $2^{\circ}\times 2^{\circ}$ degree fields imaged with the Mosaic prime-focus imager at Kitt Peak Mayall 4 m Telescope (F1 and F2) and at Cerro Tololo Blanco 4 m Telescope (F3, F4 and F5).  Each DLS field is divided into a $3\times3$ grid of $40'\times 40'$ subfields. The observing strategy was to observe in $R$ band when the seeing FWHM was $<$ 0.9$^{''}$ and in $BVz$ otherwise. Thus, the R-band imaging has fairly uniform good resolution, which is particularly convenient for weak lensing (WL) purposes. The final exposure time was 18 kiloseconds for the R band and 12 kiloseconds for the $B$,$V$ and $z$ band.

This survey has proved to be one of the deepest surveys with wide area ($>5$ deg$^2$) in the literature. It achieves 50\% completeness at $R=25.75$ (Vega) for a typical field.  The photometric redshifts have been calculated using BPZ \citep{benitez00} with tweaked templates and priors (see \citealt{schmidt13} for full details). BPZ provides a Bayesian photometric redshift, $z_b$, and a spectral type associated with this photometric redshift, $t_b$. We use the $z_b$ point estimator obtaining a photometric redshift dispersion of $\Delta z/(1+z) < 0.08$  as judged by comparison with the overlapping PRIMUS \citep{coil11} R$<$23 spectroscopic sample \citep{schmidt13}. The photometric redshift distribution of the sample peaks at $\sim$ 0.6. More details on the DLS survey can be found in Wittman et al. 2013, in preparation.

For this work, we selected all the objects that are outside of the regions masked due to bright star effects and are classified as galaxies according to the $dlsqc$ statistic. This statistic is the $\chi^2$ for a point-spread function (PSF) model fitting the second central moments of the object (Wittman et al 2013 in preparation), and we required $dlsqc>5$ for three degrees of freedom. 

Several previous papers have examined the dark matter distribution in the DLS. \cite{kubo09} presented a WL reconstruction for one of the fields (F2) in the DLS. Additionally, \cite{wittman06} studied eight shear-selected galaxy clusters in the DLS and \cite{abate09} obtained mass measurements for them. In a complementary study, \cite{sehgal08} performed a comparison between the X-ray and weak lensing masses for the four top-ranked shear-selected clumps of Abell 781, one of the clusters in the DLS.  In this paper, we will use these directly measured WL masses to compare to masses inferred from optical richness using the optical richness-mass relation of \cite{dong08}. Calibrating the richness-mass relation from the DLS data itself is beyond the scope of this paper and will be the subject of a future paper.

\section{DLS optically selected clusters}
\label{DLSopt}

We detected galaxy clusters in the DLS by using an improved version of the BCF introduced in A12.  We give a brief summary here and refer the reader to A12 for more details. The BCF calculates the probability at a given redshift that there is a cluster with a determined density and luminosity profile centered on each galaxy, including different priors related to the color-magnitude relation of the cluster or the BCG magnitude-redshift relation. We performed a search in twelve fixed redshift slices, $z_s$, from 0.1 to 1.2 with a bin width of 0.1. We set the core radius to 1.5 Mpc and the luminosity function parameters to $M^*(z=0)$=-21.44 and $\alpha=-1.05$ \citep{blanton03}. We corrected the probability assigned to those galaxies lying close to the masked stars or borders of the image by a factor proportional to the missed area.

The clusters are selected as the peaks of these probability maps. The BCF algorithm was originally validated on the CARS (A12)  dataset, and the different nature of the DLS data required one substantial adjustment to the algorithm.  Low-redshift galaxies can  be saturated in the long exposures of the DLS, which leads to poor  photometric redshift quality and confusion with stars \citep{schmidt13}.  Therefore,  we identified the cluster center as the location of the probability  peak, rather than setting the cluster location to that of the most  likely BCG as in A12. We also made one minor adjustment in the  process of merging propinquitous detections into single cluster  candidates. As in A12, we merge  detections with a redshift difference of less than two bins, but we  now consider detections separated by 1.0 Mpc or more (vs. 1.5 Mpc in  A12) to be separate clusters. Since clusters are spatially correlated and this correlation increases with cosmic time, this prevents overmerging of candidates at low redshift. Note that the merging procedure has a nontrivial effect  on the cluster counts and on the comparison of catalogues by  different authors, as discussed in A12. 
  
In A12 we extracted an optical cluster catalog from the CARS data using the BCF, and we performed simulations based on the CARS data to test the completeness and purity of the results.  Here we perform 20 realizations of similar simulations tuned to the DLS.  In these simulations, we created nine clusters with different richnesses from $\Lambda_{CL}$=10 to $\Lambda_{CL}$=200 for each redshift slice. The parameter, $\Lambda_{CL}$, is a measurement of the richness of the cluster. It is equivalent to the luminosity of the cluster in units of $L^*$. We mimicked the expected photometric redshift errors for the DLS by using a Gaussian with $\sigma=0.08(1+z_c)$. We also fixed the slope and characteristic magnitude of the luminosity function and the
core radius of the density profile as in A12. We assigned colors to mock cluster members using the prescription given by \cite{baldry04}, and then added photometric measurement errors appropriate for those magnitudes in the DLS.  In addition, we embedded these clusters in a background distribution created separately. This distribution is drawn from the redshift, magnitude and color distribution of the DLS after subtracting a first iteration of cluster detections from the survey, with positions drawn by following the Rayleigh-Levy galaxy pair separation as in A12. We used different simulated backgrounds for each realization so that artificial background detections can not repeat in all realizations.  We then simulated the effects of area lost to   bright stars by applying the actual DLS masks on the  simulations. For a more complete explanation, we refer readers to A12.  These simulations include the most important sources of  noise, but do not include some secondary sources such as cluster  ellipticity and non-Gaussian photometric redshift errors.  Because no simulation includes all the sources of error present in real data, the simulation results should be interpreted as providing an upper limit on the purity and completeness.  However, we show below (Section \$3.1) that the completeness in real data is $\sim$90\%, at least in the redshift  ($z\le 0.42$) and richness range ($M>2\times 10^{14}M_{\odot}$) in which we can directly test our sample against other samples.
  
\begin{figure}
\centering
\includegraphics[clip,angle=0,width=1.0\hsize]{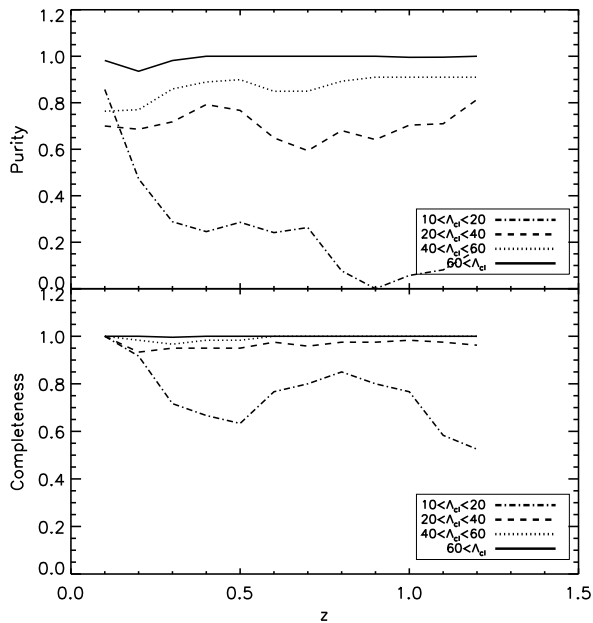} 
\caption{Completeness and purity rates for the DLS clusters as a  function of redshift and richness $\Lambda_{CL}$, based on  detections in mock data. Completeness  and purity are $\gtrsim$70\% at all redshifts for $\Lambda_{CL}\ge  40$.  Because the mock data include the most important sources of noise, performance on real data should approach these levels.}
\label{fig:completpurity}
\end{figure}

In Figure \ref{fig:completpurity}, we show these rates as a function of simulated redshift and richness. We see that the completeness and  purity are above  $\sim$ 70\% throughout the redshift range $0.1\le z\le 1.2$ for $\Lambda_{CL} \ge 40$. We also find a strong decrease of both the purity and completeness for clusters with  $\Lambda_{CL} \le$ 20.
  
We find  43.5, 53.25, 42.75, 42 and 39 clusters per square degree in Fields 1, 2, 3, 4 and 5 respectively with 0.1$\le z \le$ 1.2, detecting a total of  882 clusters down to $1.2\times 10^{14}M_{\odot}$ ($\Lambda_{CL}=40$).  We provide the complete optical cluster catalog for the DLS in Table \ref{tab:tabDLS}. The first two columns are the cluster center coordinates, the third column is the galaxy cluster redshift  estimated from the redshift slice that  maximizes the probability, and the fourth column is the galaxy cluster redshift estimated from the peak of the Gaussian fit to the galaxy population of the cluster. The fifth column refers to the $\Lambda_{CL}$ richness parameter obtained from the algorithm. The sixth column is $N_{200}$, the cluster richness considered as the number of red galaxies within $R_{200}$. The seventh column is the cluster radius, $R_{200}$, which we estimate from an empirical relation by \cite{hansen05}. Finally, we estimate the mass of the cluster, $M_{200}$, from the empirical relation between $\Lambda_{CL}$ measured from optical detections and $M_{200}$ from WL measurements by \cite{dong08}; the eighth column lists this mass estimate. The  uncertainties have been estimated by propagating the errors in the  $\Lambda_{CL}$ measurement (estimated from the simulations) into the  empirical formula by \cite{dong08}.

\begin{table*}
      \caption{Clusters detected in the DLS}
      \[
         \begin{array}{lccccccccc}
            \hline\noalign{\smallskip}
\multicolumn{1}{c}{\rm Name}&
\multicolumn{1}{c}{\rm Subfield}&
\multicolumn{1}{c}{\alpha (2000)}&
\multicolumn{1}{c}{\delta (2000)}&
\multicolumn{1}{c}{\rm z_{slice}}&
\multicolumn{1}{c}{\rm z_{est}}&
\multicolumn{1}{c}{\rm \Lambda_{CL}}&
\multicolumn{1}{c}{\rm N_{200}}&
\multicolumn{1}{c}{\rm R_{200}}&
\multicolumn{1}{c}{\rm M_{200}}\\
\multicolumn{1}{c}{}&
\multicolumn{1}{c}{}&
\multicolumn{1}{c}{($deg$)}&
\multicolumn{1}{c}{($deg$)}&
\multicolumn{1}{c}{}&
\multicolumn{1}{c}{}&
\multicolumn{1}{c}{}&
\multicolumn{1}{c}{}&
\multicolumn{1}{c}{(Mpc)}&
\multicolumn{1}{c}{(10^{14}M_{\odot})}\\
\hline\noalign{\smallskip}
\mbox{DLS0049.3+1238}  &   F1p23  & \mbox{00:49:29.06}  & \mbox{+12:38:23.64}  &        0.40  &        0.45  &   46.63  &          25  &      1.40  &            1.39^{0.52}_{0.64}  \\
\mbox{DLS0915.1+2956}  &   F2p23  & \mbox{09:15:05.73}  & \mbox{+29:56:21.84}  &        0.10  &        0.17  &  152.63  &           7  &      0.94  &            4.40^{2.26}_{2.34}  \\
\mbox{DLS0514.1-4814}  &   F3p13  & \mbox{05:14:08.98}  & \mbox{-48:14:30.47}  &        0.80  &        0.83  &   41.60  &          10  &      1.53  &            1.25^{0.95}_{0.87}  \\
\mbox{DLS1048.2-0530}  &   F4p33  & \mbox{10:48:15.65}  & \mbox{-05:30:01.44}  &        0.60  &        0.57  &   63.86  &          33  &      1.66  &            1.89^{0.89}_{1.16}  \\
\mbox{DLS1355.2-1138}  &   F5p33  & \mbox{13:55:22.82}  & \mbox{-11:38:35.52}  &        0.50  &        0.52  &   59.48  &          40  &      1.51  &            1.76^{0.89}_{0.91}  \\
\hline
         \end{array}
      \]
\begin{flushleft} 
Table \ref{tab:tabDLS} is available in the online version of the article. A portion is shown for illustration.
\end{flushleft}
\label{tab:tabDLS}
   \end{table*}

\subsection{Comparison with optical detections}

We performed a systematic search of optical cluster catalogs detected in surveys that overlap with any of the DLS fields in order to compare with the optical detections found in this work. 

We found two main optical cluster catalogs overlapping with the DLS. The first survey is the Northern Sky Optical Cluster Survey (NoSOCS, \citealt{gal00}). This survey overlaps with DLS F2, and it probes up to z $\sim$ 0.26 for the overlapping area with F2. In Table \ref{tab:knownclustersNOSOSC}, we show each cluster detected in the NoSOCS survey and its DLS counterpart, if any.  The first column refers to the name of the NoSOCS cluster, the second and third column are the cluster center coordinates and fourth column is the estimated redshift. The fifth and sixth column are the coordinates of the DLS counterpart, the seventh refers to the estimated photometric redshift, the eighth and ninth represent the angular and spatial offset between the NoSOCS and its counterpart, respectively. The physical offset, expressed in Mpc is defined here and throughout the paper as the minimum of the two physical offsets obtained for the two different redshift estimates. Finally, the tenth column represents the estimated redshift difference between both detections. We see that  all the eleven clusters detected in NoSOCS are recovered by the DLS with a redshift difference of less than $0.08(1+z_{\rm NoSOCS})$.  One of the best known galaxy clusters in the DLS is Abell 781 at $z_{spec}=0.298$ \citep{struble99}. This cluster is  detected at high S/N and recovered with redshift $z_c=0.25$, in agreement within the errors with the two clumps that NoSOCS detects at 0.2578 and 0.2575. Note that NoSOCS provides redshift estimates based on photometric redshifts with similar uncertainties as the DLS.

\begin{table*}
      \caption{Clusters detected in NoSOCS and their optically detected counterparts}
      \[
         \begin{array}{llllllllll}
            \hline\noalign{\smallskip}
\multicolumn{1}{c}{\rm Name}&
\multicolumn{1}{c}{\alpha (2000)}&
\multicolumn{1}{c}{\delta (2000)}&
\multicolumn{1}{c}{\rm z_{N}}&
\multicolumn{1}{c}{\alpha (2000)_{DLS}}&
\multicolumn{1}{c}{\delta (2000)_{DLS}}&
\multicolumn{1}{c}{\rm z_{DLS}}&
\multicolumn{1}{c}{\rm Offset}&
\multicolumn{1}{c}{\rm Offset}&
\multicolumn{1}{c}{\rm z_{N}-z_{DLS}}\\
\multicolumn{1}{c}{}&
\multicolumn{1}{c}{}&
\multicolumn{1}{c}{}&
\multicolumn{1}{c}{}&
\multicolumn{1}{c}{}&
\multicolumn{1}{c}{}&
\multicolumn{1}{c}{}&
\multicolumn{1}{c}{\rm '}&
\multicolumn{1}{c}{\rm Mpc}&
\multicolumn{1}{c}{\rm }\\
\hline\noalign{\smallskip}
\mbox{NSC091537+301312}  & \mbox{09:15:37.56}  & \mbox{+30:13:11.78}  &      0.1621  & \mbox{09:15:29.66}  & \mbox{+30:15:34.92}  &        0.23  &            2.93  &        0.49  &       -0.07  \\
\mbox{NSC091638+291943}  & \mbox{09:16:38.31}  & \mbox{+29:19:43.39}  &      0.2353  & \mbox{09:16:37.25}  & \mbox{+29:21:06.12}  &        0.21  &            1.40  &        0.29  &        0.02  \\
\mbox{NSC091715+300452}  & \mbox{09:17:15.19}  & \mbox{+30:04:51.78}  &      0.1743  & \mbox{09:17:29.86}  & \mbox{+30:04:13.80}  &        0.17  &            3.24  &        0.56  &        0.00  \\
\mbox{NSC091810+302323}  & \mbox{09:18:10.49}  & \mbox{+30:23:22.52}  &      0.1223  & \mbox{09:18:07.15}  & \mbox{+30:23:03.84}  &        0.21  &            0.78  &        0.10  &       -0.09  \\
\mbox{NSC091818+295547}  & \mbox{09:18:18.84}  & \mbox{+29:55:46.88}  &      0.1574  & \mbox{09:18:12.50}  & \mbox{+29:58:18.48}  &        0.23  &            2.88  &        0.46  &       -0.07  \\
\mbox{NSC091904+301755}  & \mbox{09:19:04.27}  & \mbox{+30:17:54.78}  &      0.1439  & \mbox{09:19:03.65}  & \mbox{+30:15:51.84}  &        0.23  &            2.05  &        0.31  &       -0.09  \\
\mbox{NSC092017+303027*}  & \mbox{09:20:17.25}  & \mbox{+30:30:26.50}  &      0.2578  & \mbox{09:20:22.27}  & \mbox{+30:29:43.44}  &        0.25  &            1.30  &        0.30  &        0.01  \\
\mbox{NSC092056+302823*}  & \mbox{09:20:56.16}  & \mbox{+30:28:22.80}  &      0.2575  & \mbox{09:20:52.89}  & \mbox{+30:28:46.92}  &        0.25  &            0.81  &        0.19  &        0.00  \\
\mbox{NSC092140+294338}  & \mbox{09:21:40.80}  & \mbox{+29:43:37.67}  &      0.2474  & \mbox{09:21:47.38}  & \mbox{+29:42:42.12}  &        0.23  &            1.70  &        0.37  &        0.02  \\
\mbox{NSC092214+310110}  & \mbox{09:22:14.45}  & \mbox{+31:01:10.20}  &      0.0541  & \mbox{09:22:33.74}  & \mbox{+30:57:51.84}  &        0.17  &            5.29  &        0.33  &       -0.11  \\
\mbox{NSC092343+304424}  & \mbox{09:23:43.75}  & \mbox{+30:44:23.93}  &      0.0862  & \mbox{09:24:00.70}  & \mbox{+30:41:52.80}  &        0.21  &            4.43  &        0.42  &       -0.13  \\
\hline
         \end{array}
      \]
\begin{flushleft} 
$^*$: These clusters correspond to different clumps of Abell 781.
\end{flushleft}
\label{tab:knownclustersNOSOSC}
   \end{table*}

There are five Abell \citep{abell89} clusters in the DLS, listed  in Table \ref{tab:knownclustersABELL}.  The BCF algorithm was  successful in identifying Abell 781 and Abell 3338 and placing them at the correct redshift.\footnote{\cite{wittman06} found a spectroscopic redshift of 0.21 for Abell 3338, superseding previous estimates which may be found in the literature.}  However, the three Abell clusters at $z<0.1$ were placed at somewhat  higher redshift (eg, Abell 1836 at $z=0.0363$ was placed at  $z=0.18$). This is due to a limitation of the data rather than the  algorithm; as explained above, some low-redshift galaxies are saturated due to the long exposures of the DLS.  These galaxies are thus eliminated from the catalog by cuts designed to eliminate  stars, or have poor photometric redshift estimates.  The NoSOCS  clusters at $z\le 0.17$ also appear at higher redshift in our  catalog for the same reason.  Therefore, users of our cluster  catalog should exercise caution when interpreting the results for  clusters with true redshifts $<0.2$ or with redshifts listed in our  catalog as $<0.25$.

\begin{table*}
      \caption{Known Abell clusters and their optically detected counterparts.}
      \[
         \begin{array}{llllllllll}
            \hline\noalign{\smallskip}
\multicolumn{1}{c}{\rm Name}&
\multicolumn{1}{c}{\alpha (2000)}&
\multicolumn{1}{c}{\delta (2000)}&
\multicolumn{1}{c}{\rm z_{spec}}&
\multicolumn{1}{c}{\alpha (2000)_{DLS}}&
\multicolumn{1}{c}{\delta (2000)_{DLS}}&
\multicolumn{1}{c}{\rm z_{DLS}}&
\multicolumn{1}{c}{\rm Offset}&
\multicolumn{1}{c}{\rm Offset}&
\multicolumn{1}{c}{\rm z_{Ab}-z_{DLS}}\\
\multicolumn{1}{c}{}&
\multicolumn{1}{c}{}&
\multicolumn{1}{c}{}&
\multicolumn{1}{c}{}&
\multicolumn{1}{c}{}&
\multicolumn{1}{c}{}&
\multicolumn{1}{c}{}&
\multicolumn{1}{c}{\rm '}&
\multicolumn{1}{c}{\rm Mpc}&
\multicolumn{1}{c}{\rm }\\
\hline\noalign{\smallskip}
\mbox{ABELL0781}  & \mbox{09:20:23.18}  & \mbox{+30:26:15.00}  &      0.2980  & \mbox{09:20:32.35}  & \mbox{+30:25:11.64}  &        0.25  &            2.24  &        0.53  &        0.04  \\
\mbox{ABELL3330}  & \mbox{05:14:40.01}  & \mbox{-49:03:14.99}  &      0.0921  & \mbox{05:14:32.35}  & \mbox{-49:01:30.36}  &        0.22  &            2.15  &        0.22  &       -0.13  \\
\mbox{ABELL3338}  & \mbox{05:22:37.80}  & \mbox{-48:16:13.99}  &      0.2100  & \mbox{05:22:47.78}  & \mbox{-48:18:05.77}  &        0.23  &            2.50  &        0.51  &       -0.02  \\
\mbox{ABELL1836}  & \mbox{14:01:40.61}  & \mbox{-11:36:27.00}  &      0.0363  & \mbox{14:01:50.07}  & \mbox{-11:38:07.80}  &        0.18  &            2.86  &        0.12  &       -0.14  \\
\mbox{ABELL1837}  & \mbox{14:01:46.30}  & \mbox{-11:09:27.00}  &      0.0700  & \mbox{14:01:34.06}  & \mbox{-11:08:12.48}  &        0.22  &            3.25  &        0.26  &       -0.15  \\
\hline
         \end{array}
      \]
\begin{flushleft} 
\end{flushleft}
\label{tab:knownclustersABELL}
   \end{table*}

The second optical survey we compared our detections with is the SDSS, which has many catalogs of clusters already available. We choose for comparison the catalog by \cite{wen12} since it is based on the DR8 SDSS-III data and overlaps with the largest amount of data in the DLS (three fields, F1, F2 and F5).

In Tables \ref{tab:knownclustersSDSSF1}, \ref{tab:knownclustersSDSSF2} and \ref{tab:knownclustersSDSSF5}, we show each cluster detected in the SDSS survey and its DLS counterpart in F1, F2 and F5 respectively, if detected. If the best match was found at distance larger than 1.5 Mpc or has a redshift difference larger than $2(1+z_c)0.08$, this is not considered a match. We find that  89.36\% of the clusters in the area of SDSS that overlaps with the DLS are detected in the DLS. For this matching, the mean and $rms$ redshift differences are  -0.002 and 0.070, respectively.  The  recovery rate of SDSS clusters provides further evidence that the DLS sample completeness is $\sim$90\% at least up to z$\sim$ 0.42. Conversely, we find that  50.58\% of the clusters in the three fields of the DLS that overlap with the SDSS within 0.1$\le z\le $0.42 are found in the SDSS. This redshift upper limit corresponds to the redshift at which the completeness of the SDSS cluster sample is $>$95\%. The mean and $rms$ redshift differences are -0.03 and  0.009, respectively. Note that the DLS is deeper and therefore complete to a lower limit than the SDSS so we do expect some real clusters not to be found in the SDSS.

\renewcommand{\tabcolsep}{0.1pt}
\begin{table*}
      \caption{Clusters detected in the SDSS and their optically detected counterparts in F1}
      \[
         \begin{array}{lllllllllr}
            \hline\noalign{\smallskip}
\multicolumn{1}{c}{\rm Name}&
\multicolumn{1}{c}{\alpha (2000)}&
\multicolumn{1}{c}{\delta (2000)}&
\multicolumn{1}{c}{\rm z_{W}}&
\multicolumn{1}{c}{\alpha (2000)_{A13}}&
\multicolumn{1}{c}{\delta (2000)_{A13}}&
\multicolumn{1}{c}{\rm z_{A13}}&
\multicolumn{1}{c}{\rm Offset}&
\multicolumn{1}{c}{\rm Offset}&
\multicolumn{1}{c}{\rm z_{W}-z_{A13}}\\
\multicolumn{1}{c}{}&
\multicolumn{1}{c}{}&
\multicolumn{1}{c}{}&
\multicolumn{1}{c}{}&
\multicolumn{1}{c}{}&
\multicolumn{1}{c}{}&
\multicolumn{1}{c}{}&
\multicolumn{1}{c}{\rm '}&
\multicolumn{1}{c}{\rm Mpc}&
\multicolumn{1}{c}{\rm }\\
\hline\noalign{\smallskip}
\mbox{WHLJ005049.7+121613}  & \mbox{00:50:49.70}  & \mbox{+12:16:12.72}  &      0.5483  & \mbox{00:50:41.47}  & \mbox{+12:17:30.12}  &        0.56  &            2.39  &        0.92  &       -0.01  \\
\mbox{WHLJ005107.6+130214}  & \mbox{00:51:07.61}  & \mbox{+13:02:13.92}  &      0.4545  & \mbox{00:51:07.61}  & \mbox{+13:02:14.28}  &        0.50  &            0.01  &        0.00  &       -0.05  \\
\mbox{WHLJ005129.8+125937}  & \mbox{00:51:29.78}  & \mbox{+12:59:37.32}  &      0.6337  & \mbox{00:51:29.78}  & \mbox{+12:59:37.32}  &        0.60  &            0.00  &        0.00  &        0.04  \\
\mbox{WHLJ005150.6+123456}  & \mbox{00:51:50.59}  & \mbox{+12:34:56.28}  &      0.1893  & \mbox{00:51:50.59}  & \mbox{+12:34:56.64}  &        0.21  &            0.01  &        0.00  &       -0.02  \\
\mbox{WHLJ005215.0+115011}  & \mbox{00:52:15.00}  & \mbox{+11:50:11.04}  &      0.4403  & \mbox{00:52:14.98}  & \mbox{+11:50:11.40}  &        0.50  &            0.01  &        0.00  &       -0.06  \\
\mbox{WHLJ005228.2+122052}  & \mbox{00:52:28.20}  & \mbox{+12:20:52.08}  &      0.4371  & \mbox{00:52:28.20}  & \mbox{+12:20:52.08}  &        0.47  &            0.00  &        0.00  &       -0.04  \\
\mbox{WHLJ005239.5+121857}  & \mbox{00:52:39.50}  & \mbox{+12:18:56.88}  &      0.4259  & \mbox{00:52:39.48}  & \mbox{+12:18:57.24}  &        0.48  &            0.01  &        0.00  &       -0.05  \\
\mbox{WHLJ005311.6+122339}  & \mbox{00:53:11.57}  & \mbox{+12:23:39.12}  &      0.3812  & \mbox{00:53:11.59}  & \mbox{+12:23:39.48}  &        0.45  &            0.01  &        0.00  &       -0.07  \\
\mbox{WHLJ005328.6+123031}  & \mbox{00:53:28.61}  & \mbox{+12:30:30.96}  &      0.4224  & \mbox{00:53:28.34}  & \mbox{+12:29:13.92}  &        0.47  &            1.29  &        0.43  &       -0.05  \\
\mbox{WHLJ005359.9+121905}  & \mbox{00:53:59.95}  & \mbox{+12:19:04.80}  &      0.3908  & \mbox{00:54:06.02}  & \mbox{+12:18:56.88}  &        0.55  &            1.49  &        0.47  &       -0.16  \\
\mbox{WHLJ005402.2+130324}  & \mbox{00:54:02.21}  & \mbox{+13:03:23.76}  &      0.3542  & \mbox{00:54:00.07}  & \mbox{+13:03:16.20}  &        0.25  &            0.54  &        0.12  &        0.10  \\
\mbox{WHLJ005402.4+125912}  & \mbox{00:54:02.45}  & \mbox{+12:59:12.48}  &      0.3599  & \mbox{00:54:02.14}  & \mbox{+12:59:13.20}  &        0.36  &            0.08  &        0.02  &        0.00  \\
\mbox{WHLJ005413.3+114731}  & \mbox{00:54:13.34}  & \mbox{+11:47:30.84}  &      0.5982  & \mbox{00:54:14.74}  & \mbox{+11:47:22.20}  &        0.54  &            0.37  &        0.14  &        0.06  \\
\mbox{WHLJ005423.8+123755}  & \mbox{00:54:23.81}  & \mbox{+12:37:54.48}  &      0.3325  & \mbox{00:54:23.81}  & \mbox{+12:37:54.84}  &        0.34  &            0.01  &        0.00  &       -0.01  \\
\mbox{WHLJ005430.7+123306}  & \mbox{00:54:30.72}  & \mbox{+12:33:06.12}  &      0.3395  & \mbox{00:54:30.79}  & \mbox{+12:33:14.76}  &        0.39  &            0.15  &        0.04  &       -0.05  \\
\mbox{WHLJ005500.5+125803}  & \mbox{00:55:00.55}  & \mbox{+12:58:03.00}  &      0.3894  & \mbox{00:55:00.94}  & \mbox{+12:59:08.52}  &        0.45  &            1.10  &        0.35  &       -0.06  \\
\mbox{WHLJ005534.0+114632}  & \mbox{00:55:33.96}  & \mbox{+11:46:31.62}  &      0.2997  & -      & -  & - & - & - & -  \\
\mbox{WHLJ005555.0+114759}  & \mbox{00:55:54.96}  & \mbox{+11:47:59.64}  &      0.5362  & \mbox{00:55:54.94}  & \mbox{+11:48:00.00}  &        0.53  &            0.01  &        0.00  &        0.00  \\
\mbox{WHLJ005559.4+121141}  & \mbox{00:55:59.40}  & \mbox{+12:11:41.28}  &      0.5673  & \mbox{00:55:55.39}  & \mbox{+12:10:37.56}  &        0.52  &            1.44  &        0.54  &        0.04  \\
\mbox{WHLJ005601.5+131028}  & \mbox{00:56:01.51}  & \mbox{+13:10:27.48}  &      0.3164  & \mbox{00:55:56.78}  & \mbox{+13:11:48.12}  &        0.25  &            1.77  &        0.41  &        0.07  \\ 
\mbox{WHLJ005615.6+120329}  & \mbox{00:56:15.58}  & \mbox{+12:03:28.80}  &      0.3698  & \mbox{00:56:15.58}  & \mbox{+12:03:29.16}  &        0.51  &            0.01  &        0.00  &       -0.14  \\
\mbox{WHLJ005631.6+120057}  & \mbox{00:56:31.58}  & \mbox{+12:00:57.24}  &      0.5133  & \mbox{00:56:31.54}  & \mbox{+12:00:58.68}  &        0.50  &            0.03  &        0.01  &        0.01  \\
 \hline
         \end{array}
      \]
\label{tab:knownclustersSDSSF1}
   \end{table*}

\begin{table*}
      \caption{Clusters detected in the SDSS and their optically detected counterparts in F2}
      \[
         \begin{array}{lllllllllr}
            \hline\noalign{\smallskip}
\multicolumn{1}{c}{\rm Name}&
\multicolumn{1}{c}{\alpha (2000)}&
\multicolumn{1}{c}{\delta (2000)}&
\multicolumn{1}{c}{\rm z_{W}}&
\multicolumn{1}{c}{\alpha (2000)_{A13}}&
\multicolumn{1}{c}{\delta (2000)_{A13}}&
\multicolumn{1}{c}{\rm z_{A13}}&
\multicolumn{1}{c}{\rm Offset}&
\multicolumn{1}{c}{\rm Offset}&
\multicolumn{1}{c}{\rm z_{W}-z_{A13}}\\
\multicolumn{1}{c}{}&
\multicolumn{1}{c}{}&
\multicolumn{1}{c}{}&
\multicolumn{1}{c}{}&
\multicolumn{1}{c}{}&
\multicolumn{1}{c}{}&
\multicolumn{1}{c}{}&
\multicolumn{1}{c}{\rm '}&
\multicolumn{1}{c}{\rm Mpc}&
\multicolumn{1}{c}{\rm }\\
\hline\noalign{\smallskip}
\mbox{WHLJ091557.2+301122}  & \mbox{09:15:57.17}  & \mbox{+30:11:22.20}  &      0.4203  & -      & -  & - & - & - & -  \\ 
\mbox{WHLJ091559.7+292530}  & \mbox{09:15:59.71}  & \mbox{+29:25:30.36}  &      0.5438  & \mbox{09:16:09.38}  & \mbox{+29:25:44.76}  &        0.50  &            2.12  &        0.77  &        0.04  \\ 
\mbox{WHLJ091608.2+295232}  & \mbox{09:16:08.16}  & \mbox{+29:52:32.52}  &      0.6412  & \mbox{09:16:10.44}  & \mbox{+29:52:17.76}  &        0.52  &            0.55  &        0.21  &        0.12  \\
\mbox{WHLJ091608.5+292839}  & \mbox{09:16:08.50}  & \mbox{+29:28:39.36}  &      0.5423  & \mbox{09:16:08.52}  & \mbox{+29:28:39.72}  &        0.50  &            0.01  &        0.00  &        0.04  \\
\mbox{WHLJ091622.9+291620}  & \mbox{09:16:22.87}  & \mbox{+29:16:20.64}  &      0.5625  & \mbox{09:16:18.31}  & \mbox{+29:17:40.92}  &        0.52  &            1.67  &        0.62  &        0.05  \\ 
\mbox{WHLJ091625.8+295208}  & \mbox{09:16:25.80}  & \mbox{+29:52:07.68}  &      0.5262  & \mbox{09:16:10.44}  & \mbox{+29:52:17.76}  &        0.52  &            3.33  &        1.25  &        0.00  \\ 
\mbox{WHLJ091636.0+300237}  & \mbox{09:16:36.05}  & \mbox{+30:02:36.60}  &      0.5140  & \mbox{09:16:40.05}  & \mbox{+30:03:46.44}  &        0.51  &            1.45  &        0.53  &        0.01  \\
\mbox{WHLJ091639.7+304218}  & \mbox{09:16:39.72}  & \mbox{+30:42:17.64}  &      0.4743  & \mbox{09:16:39.72}  & \mbox{+30:42:17.64}  &        0.50  &            0.00  &        0.00  &       -0.03  \\
\mbox{WHLJ091650.2+305239}  & \mbox{09:16:50.19}  & \mbox{+30:52:39.36}  &      0.5494  & \mbox{09:16:54.14}  & \mbox{+30:51:09.72}  &        0.59  &            1.72  &        0.66  &       -0.04  \\
\mbox{WHLJ091655.5+300015}  & \mbox{09:16:55.44}  & \mbox{+30:00:14.76}  &      0.5488  & \mbox{09:16:45.70}  & \mbox{+30:01:11.28}  &        0.51  &            2.31  &        0.85  &        0.04  \\ 
\mbox{WHLJ091705.9+300118}  & \mbox{09:17:05.93}  & \mbox{+30:01:18.48}  &      0.3128  & \mbox{09:17:03.26}  & \mbox{+30:01:20.64}  &        0.24  &            0.58  &        0.13  &        0.07  \\
\mbox{WHLJ091714.5+301737}  & \mbox{09:17:14.49}  & \mbox{+30:17:36.96}  &      0.5380  & \mbox{09:17:11.71}  & \mbox{+30:16:47.28}  &        0.53  &            1.02  &        0.38  &        0.01  \\
\mbox{WHLJ091717.9+301009}  & \mbox{09:17:17.90}  & \mbox{+30:10:08.40}  &      0.2144  & \mbox{09:17:17.93}  & \mbox{+30:10:08.76}  &        0.22  &            0.01  &        0.00  &       -0.00  \\
\mbox{WHLJ091729.9+300414}  & \mbox{09:17:29.86}  & \mbox{+30:04:13.44}  &      0.1634  & \mbox{09:17:29.86}  & \mbox{+30:04:13.80}  &        0.17  &            0.01  &        0.00  &       -0.01  \\
\mbox{WHLJ091740.6+295523}  & \mbox{09:17:40.58}  & \mbox{+29:55:22.80}  &      0.4821  & -      & -  & - & - & - & -  \\ 
\mbox{WHLJ091812.5+295818}  & \mbox{09:18:12.48}  & \mbox{+29:58:18.12}  &      0.1710  & \mbox{09:18:12.50}  & \mbox{+29:58:18.48}  &        0.23  &            0.01  &        0.00  &       -0.06  \\
\mbox{WHLJ091820.1+302416}  & \mbox{09:18:20.09}  & \mbox{+30:24:16.20}  &      0.1325  & \mbox{09:18:07.15}  & \mbox{+30:23:03.84}  &        0.21  &            3.04  &        0.42  &       -0.08  \\
\mbox{WHLJ091826.6+293412}  & \mbox{09:18:26.62}  & \mbox{+29:34:12.00}  &      0.4484  & \mbox{09:18:32.07}  & \mbox{+29:32:45.60}  &        0.63  &            1.86  &        0.64  &       -0.18  \\
\mbox{WHLJ091836.1+295308}  & \mbox{09:18:36.05}  & \mbox{+29:53:07.80}  &      0.3164  & \mbox{09:18:34.34}  & \mbox{+29:53:18.60}  &        0.27  &            0.41  &        0.10  &        0.05  \\
\mbox{WHLJ091836.4+292655}  & \mbox{09:18:36.41}  & \mbox{+29:26:54.96}  &      0.5715  & \mbox{09:18:36.43}  & \mbox{+29:26:54.96}  &        0.51  &            0.00  &        0.00  &        0.06  \\
\mbox{WHLJ091839.8+300625}  & \mbox{09:18:39.77}  & \mbox{+30:06:25.56}  &      0.1373  & \mbox{09:18:12.50}  & \mbox{+29:58:18.48}  &        0.23  &           10.04  &        1.45  &       -0.09  \\ 
\mbox{WHLJ091856.2+302059}  & \mbox{09:18:56.16}  & \mbox{+30:20:59.28}  &      0.2846  & \mbox{09:18:55.97}  & \mbox{+30:22:10.20}  &        0.48  &            1.18  &        0.30  &       -0.19  \\ 
\mbox{WHLJ091928.9+292011}  & \mbox{09:19:28.95}  & \mbox{+29:20:10.68}  &      0.3484  & \mbox{09:19:21.38}  & \mbox{+29:16:46.56}  &        0.47  &            3.78  &        1.11  &       -0.12  \\ 
\mbox{WHLJ091935.0+303156}  & \mbox{09:19:35.04}  & \mbox{+30:31:56.28}  &      0.4310  & \mbox{09:19:34.27}  & \mbox{+30:30:28.44}  &        0.49  &            1.47  &        0.49  &       -0.06  \\
\mbox{WHLJ091936.9+293910}  & \mbox{09:19:36.91}  & \mbox{+29:39:10.08}  &      0.3003  &  -      & -  & - & - & - & -  \\ 
\mbox{WHLJ091938.7+303256}  & \mbox{09:19:38.71}  & \mbox{+30:32:55.68}  &      0.5415  & \mbox{09:19:34.27}  & \mbox{+30:30:28.44}  &        0.49  &            2.63  &        0.95  &        0.05  \\
\mbox{WHLJ092007.2+302934}  & \mbox{09:20:07.25}  & \mbox{+30:29:34.08}  &      0.4310  & \mbox{09:20:22.27}  & \mbox{+30:29:43.44}  &        0.25  &            3.24  &        0.76  &        0.18  \\
\mbox{WHLJ092025.8+302939}  & \mbox{09:20:25.80}  & \mbox{+30:29:38.76}  &      0.2928  & \mbox{09:20:22.27}  & \mbox{+30:29:43.44}  &        0.25  &            0.76  &        0.18  &        0.04  \\
\mbox{WHLJ092031.3+291139}  & \mbox{09:20:31.25}  & \mbox{+29:11:39.48}  &      0.2042  & \mbox{09:20:31.27}  & \mbox{+29:11:39.84}  &        0.23  &            0.01  &        0.00  &       -0.03  \\
\mbox{WHLJ092037.6+302555}  & \mbox{09:20:37.59}  & \mbox{+30:25:54.84}  &      0.4311  & \mbox{09:20:32.35}  & \mbox{+30:25:11.64}  &        0.25  &            1.34  &        0.32  &        0.18  \\
\mbox{WHLJ092052.8+294113}  & \mbox{09:20:52.82}  & \mbox{+29:41:12.84}  &      0.2805  &  -      & -  & - & - & - & -  \\ 
\mbox{WHLJ092055.4+300506}  & \mbox{09:20:55.41}  & \mbox{+30:05:06.00}  &      0.5430  & \mbox{09:20:54.70}  & \mbox{+30:04:09.84}  &        0.55  &            0.95  &        0.36  &       -0.00  \\
\mbox{WHLJ092056.4+294703}  & \mbox{09:20:56.43}  & \mbox{+29:47:03.48}  &      0.4169  & \mbox{09:20:50.69}  & \mbox{+29:46:01.20}  &        0.41  &            1.62  &        0.53  &        0.01  \\ 
\mbox{WHLJ092102.3+303119}  & \mbox{09:21:02.26}  & \mbox{+30:31:19.56}  &      0.3129  & \mbox{09:20:52.89}  & \mbox{+30:28:46.92}  &        0.25  &            3.25  &        0.76  &        0.06  \\ 
\mbox{WHLJ092110.8+302806}  & \mbox{09:21:10.83}  & \mbox{+30:28:06.24}  &      0.4180  & \mbox{09:21:11.18}  & \mbox{+30:27:45.00}  &        0.41  &            0.36  &        0.12  &        0.01  \\
\mbox{WHLJ092113.2+301222}  & \mbox{09:21:13.20}  & \mbox{+30:12:21.60}  &      0.6881  & \mbox{09:21:23.49}  & \mbox{+30:13:04.80}  &        0.54  &            2.34  &        0.89  &        0.15  \\
\mbox{WHLJ092116.2+303030}  & \mbox{09:21:16.22}  & \mbox{+30:30:30.24}  &      0.5439  & \mbox{09:21:22.13}  & \mbox{+30:29:10.68}  &        0.47  &            1.84  &        0.65  &        0.07  \\
\mbox{WHLJ092121.0+301334}  & \mbox{09:21:21.00}  & \mbox{+30:13:34.32}  &      0.3140  & \mbox{09:21:23.49}  & \mbox{+30:13:04.80}  &        0.54  &            0.73  &        0.20  &       -0.22  \\
\mbox{WHLJ092129.3+295735}  & \mbox{09:21:29.32}  & \mbox{+29:57:35.28}  &      0.4196  & \mbox{09:21:38.57}  & \mbox{+29:58:03.00}  &        0.46  &            2.05  &        0.68  &       -0.04  \\
\mbox{WHLJ092136.3+292730}  & \mbox{09:21:36.34}  & \mbox{+29:27:30.24}  &      0.3738  & \mbox{09:21:36.02}  & \mbox{+29:27:30.24}  &        0.23  &            0.07  &        0.02  &        0.15  \\
\mbox{WHLJ092139.1+301929}  & \mbox{09:21:39.07}  & \mbox{+30:19:28.56}  &      0.5614  & \mbox{09:21:36.12}  & \mbox{+30:17:51.00}  &        0.51  &            1.75  &        0.65  &        0.05  \\
\mbox{WHLJ092139.7+294545}  & \mbox{09:21:39.72}  & \mbox{+29:45:44.64}  &      0.2533  & \mbox{09:21:39.14}  & \mbox{+29:46:14.52}  &        0.22  &            0.51  &        0.11  &        0.03  \\ 
\mbox{WHLJ092156.0+292613}  & \mbox{09:21:55.99}  & \mbox{+29:26:13.56}  &      0.3935  & \mbox{09:22:01.85}  & \mbox{+29:25:50.52}  &        0.50  &            1.33  &        0.42  &       -0.11  \\
\mbox{WHLJ092218.6+295630}  & \mbox{09:22:18.60}  & \mbox{+29:56:30.12}  &      0.5012  & \mbox{09:22:18.60}  & \mbox{+29:56:30.12}  &        0.52  &            0.00  &        0.00  &       -0.02  \\
\mbox{WHLJ092223.2+301803}  & \mbox{09:22:23.16}  & \mbox{+30:18:02.52}  &      0.4737  & \mbox{09:22:20.38}  & \mbox{+30:17:02.40}  &        0.50  &            1.17  &        0.41  &       -0.03  \\
\mbox{WHLJ092241.6+300713}  & \mbox{09:22:41.59}  & \mbox{+30:07:13.08}  &      0.4177  & \mbox{09:22:40.82}  & \mbox{+30:07:25.32}  &        0.48  &            0.26  &        0.09  &       -0.06  \\
\mbox{WHLJ092255.4+304727}  & \mbox{09:22:55.39}  & \mbox{+30:47:26.88}  &      0.6484  &  -      & -  & - & - & - & -  \\ 
\mbox{WHLJ092304.8+295909}  & \mbox{09:23:04.80}  & \mbox{+29:59:08.88}  &      0.3125  & \mbox{09:23:07.20}  & \mbox{+30:00:54.36}  &        0.50  &            1.83  &        0.50  &       -0.19  \\
\mbox{WHLJ092317.9+304446}  & \mbox{09:23:17.95}  & \mbox{+30:44:46.32}  &      0.5133  & \mbox{09:23:28.92}  & \mbox{+30:44:53.16}  &        0.47  &            2.36  &        0.83  &        0.05  \\ 
 \hline
         \end{array}
      \]
\label{tab:knownclustersSDSSF2}
   \end{table*}

\begin{table*}
      \caption{Clusters detected in the SDSS and their optically detected counterparts in F5}
      \[
         \begin{array}{lllllllllr}
            \hline\noalign{\smallskip}
\multicolumn{1}{c}{\rm Name}&
\multicolumn{1}{c}{\alpha (2000)}&
\multicolumn{1}{c}{\delta (2000)}&
\multicolumn{1}{c}{\rm z_{W}}&
\multicolumn{1}{c}{\alpha (2000)_{A13}}&
\multicolumn{1}{c}{\delta (2000)_{A13}}&
\multicolumn{1}{c}{\rm z_{A13}}&
\multicolumn{1}{c}{\rm Offset}&
\multicolumn{1}{c}{\rm Offset}&
\multicolumn{1}{c}{\rm z_{W}-z_{A13}}\\
\multicolumn{1}{c}{}&
\multicolumn{1}{c}{}&
\multicolumn{1}{c}{}&
\multicolumn{1}{c}{}&
\multicolumn{1}{c}{}&
\multicolumn{1}{c}{}&
\multicolumn{1}{c}{}&
\multicolumn{1}{c}{\rm '}&
\multicolumn{1}{c}{\rm Mpc}&
\multicolumn{1}{c}{\rm }\\
\hline\noalign{\smallskip}
\mbox{WHLJ135732.0-103021}  & \mbox{13:57:32.02}  & \mbox{-10:30:20.52}  &      0.2094  & -      & -  & - & - & - & -  \\
\mbox{WHLJ135754.3-102818}  & \mbox{13:57:54.29}  & \mbox{-10:28:18.12}  &      0.2222  & -      & -  & - & - & - & -  \\
\mbox{WHLJ135805.2-104630}  & \mbox{13:58:05.23}  & \mbox{-10:46:29.64}  &      0.4870  & -      & -  & - & - & - & -  \\
\mbox{WHLJ135810.9-105248}  & \mbox{13:58:10.92}  & \mbox{-10:52:48.36}  &      0.4778  & \mbox{13:58:10.92}  & \mbox{-10:52:48.00}  &        0.40  &            0.01  &        0.00  &        0.07  \\
\mbox{WHLJ135842.5-104111}  & \mbox{13:58:42.45}  & \mbox{-10:41:11.40}  &      0.5028  & \mbox{13:58:42.48}  & \mbox{-10:41:11.04}  &        0.49  &            0.01  &        0.00  &        0.01  \\
\mbox{WHLJ135854.2-110106}  & \mbox{13:58:54.17}  & \mbox{-11:01:05.88}  &      0.1780  & \mbox{13:58:58.08}  & \mbox{-11:00:02.52}  &        0.17  &            1.43  &        0.24  &        0.01  \\
\mbox{WHLJ135858.1-102626}  & \mbox{13:58:58.10}  & \mbox{-10:26:25.80}  &      0.3887  & -      & -  & - & - & - & -  \\
\mbox{WHLJ135943.5-113936}  & \mbox{13:59:43.44}  & \mbox{-11:39:36.00}  &      0.2720  & \mbox{13:59:43.47}  & \mbox{-11:39:36.00}  &        0.42  &            0.01  &        0.00  &       -0.15  \\
\mbox{WHLJ135952.8-101543}  & \mbox{13:59:52.77}  & \mbox{-10:15:43.20}  &      0.3927  & \mbox{13:59:57.84}  & \mbox{-10:15:08.64}  &        0.44  &            1.37  &        0.43  &       -0.05  \\
\mbox{WHLJ140000.8-102602}  & \mbox{14:00:00.84}  & \mbox{-10:26:01.68}  &      0.4706  & \mbox{14:00:00.87}  & \mbox{-10:26:01.32}  &        0.51  &            0.01  &        0.00  &       -0.04  \\
\mbox{WHLJ140010.1-104315}  & \mbox{14:00:10.13}  & \mbox{-10:43:15.60}  &      0.4400   & -      & -  & - & - & - & -  \\
\mbox{WHLJ140039.6-105648}  & \mbox{14:00:39.57}  & \mbox{-10:56:47.76}  &      0.4118   & -      & -  & - & - & - & -  \\
\mbox{WHLJ140110.6-113851}  & \mbox{14:01:10.63}  & \mbox{-11:38:51.36}  &      0.1730  & \mbox{14:01:03.48}  & \mbox{-11:36:22.32}  &        0.20  &            3.04  &        0.53  &       -0.03  \\
\mbox{WHLJ140127.4-113155}  & \mbox{14:01:27.41}  & \mbox{-11:31:54.84}  &      0.1794  & \mbox{14:01:27.41}  & \mbox{-11:31:54.84}  &        0.22  &            0.00  &        0.00  &       -0.04  \\
\mbox{WHLJ140133.2-101406}  & \mbox{14:01:33.17}  & \mbox{-10:14:06.00}  &      0.4603  & \mbox{14:01:38.21}  & \mbox{-10:14:08.16}  &        0.49  &            1.24  &        0.43  &       -0.03  \\
\mbox{WHLJ140136.4-110744}  & \mbox{14:01:36.38}  & \mbox{-11:07:43.68}  &      0.0692  & \mbox{14:01:34.06}  & \mbox{-11:08:12.48}  &        0.22  &            0.75  &        0.06  &       -0.15  \\
\mbox{WHLJ140142.1-104855}  & \mbox{14:01:42.10}  & \mbox{-10:48:55.08}  &      0.1663  & \mbox{14:01:42.12}  & \mbox{-10:48:55.08}  &        0.20  &            0.01  &        0.00  &       -0.03  \\
\mbox{WHLJ140145.7-103403}  & \mbox{14:01:45.74}  & \mbox{-10:34:02.64}  &      0.2635  & \mbox{14:01:45.77}  & \mbox{-10:34:02.64}  &        0.22  &            0.01  &        0.00  &        0.05  \\
\mbox{WHLJ140149.5-102602}  & \mbox{14:01:49.49}  & \mbox{-10:26:02.04}  &      0.4094  & \mbox{14:02:00.19}  & \mbox{-10:23:45.96}  &        0.42  &            3.47  &        1.13  &       -0.01  \\
\mbox{WHLJ140153.9-103756}  & \mbox{14:01:53.95}  & \mbox{-10:37:55.92}  &      0.2767  & \mbox{14:01:53.64}  & \mbox{-10:37:51.24}  &        0.22  &            0.11  &        0.02  &        0.06  \\
\mbox{WHLJ140200.5-102249}  & \mbox{14:02:00.51}  & \mbox{-10:22:49.08}  &      0.4308  & \mbox{14:02:00.19}  & \mbox{-10:23:45.96}  &        0.42  &            0.95  &        0.32  &        0.01  \\
\mbox{WHLJ140234.0-104553}  & \mbox{14:02:33.98}  & \mbox{-10:45:53.64}  &      0.2668  & \mbox{14:02:33.98}  & \mbox{-10:45:53.28}  &        0.23  &            0.01  &        0.00  &        0.04  \\
\mbox{WHLJ140248.8-110117}  & \mbox{14:02:48.77}  & \mbox{-11:01:17.40}  &      0.4001  & \mbox{14:02:44.78}  & \mbox{-11:01:05.16}  &        0.38  &            1.00  &        0.31  &        0.02  \\
  \hline
         \end{array}
      \]
\label{tab:knownclustersSDSSF5}
   \end{table*}

\subsection{Comparison with X-ray detections}

We also cross-correlated the detections found in the DLS with the detections found in any existing X-ray cluster catalog. To do this, we performed a search for X-ray detected galaxy clusters in the NASA/IPAC Extragalactic Database (NED). We found twelve galaxy clusters in the five fields of the DLS. Note that this is not an X-ray selected sample; it  consists of pointed followup of clusters detected by other means.  Thus we can test the fraction of known X-ray clusters that are detected optically, but this should not be interpreted as testing the fraction of X-ray-selected clusters that are detected optically.

In Table \ref{tab:knownclustersX}, we list the clusters detected in X-rays with their DLS optically-detected counterparts. We find that eleven out of these twelve clusters are well recovered by the DLS with a redshift difference of less than $0.08(1+z)$ and a maximum  offset of 0.83 Mpc. The only cluster that we do not detect,  CXOUJ091554+293316, happens to be centered close to a saturated  star, and therefore, most of the galaxies in the cluster are masked  out. The mean and $rms$ redshift differences for the whole X-ray sample are 0.033 and 0.047, respectively. We are not able to compare X-ray  properties with optical estimates due to the lack of homogeneous  measurements for the X-ray sample.

\begin{table*}
      \caption{Clusters detected in X-rays and their optically
        detected counterparts}
      \[
         \begin{array}{lllllllllr}
            \hline\noalign{\smallskip}
\multicolumn{1}{c}{\rm Name}&
\multicolumn{1}{c}{\alpha (2000)}&
\multicolumn{1}{c}{\delta (2000)}&
\multicolumn{1}{c}{\rm z_{X}}&
\multicolumn{1}{c}{\alpha (2000)_{A13}}&
\multicolumn{1}{c}{\delta (2000)_{A13}}&
\multicolumn{1}{c}{\rm z_{13}}&
\multicolumn{1}{c}{\rm Offset}&
\multicolumn{1}{c}{\rm Offset}&
\multicolumn{1}{c}{\rm z_{X}-z_{A13}}\\
\multicolumn{1}{c}{}&
\multicolumn{1}{c}{}&
\multicolumn{1}{c}{}&
\multicolumn{1}{c}{}&
\multicolumn{1}{c}{}&
\multicolumn{1}{c}{}&
\multicolumn{1}{c}{}&
\multicolumn{1}{c}{\rm '}&
\multicolumn{1}{c}{\rm Mpc}&
\multicolumn{1}{c}{\rm }\\
\hline\noalign{\smallskip}
\mbox{CXOUJ091551+293637}  & \mbox{09:15:51.79}  & \mbox{+29:36:37.00}  &      0.5300  & \mbox{09:15:59.26}  & \mbox{+29:37:08.04}  &        0.55  &            1.70  &        0.64  &       -0.02  \\
\mbox{CXOUJ091554+293316}  & \mbox{09:15:54.41}  & \mbox{+29:33:16.00}  &      0.1847   & -      & -  & - & - & - & -   \\
\mbox{CXOUJ091601+292750}  & \mbox{09:16:01.10}  & \mbox{+29:27:50.00}  &      0.5310  & \mbox{09:16:08.52}  & \mbox{+29:28:39.72}  &        0.50  &            1.82  &        0.66  &        0.03  \\
\mbox{*1RXSJ092025.5+30315}  & \mbox{09:20:25.49}  & \mbox{+30:31:54.00}  &      0.2952  & \mbox{09:20:21.29}  & \mbox{+30:33:42.84}  &        0.27  &            2.03  &        0.50  &        0.03  \\
\mbox{CXOUJ052147-482124}  & \mbox{05:21:47.90}  & \mbox{-48:21:24.00}  &      0.3000  & \mbox{05:21:46.44}  & \mbox{-48:21:59.76}  &        0.27  &            0.64  &        0.16  &        0.03  \\
\mbox{CXOUJ052159-481606}  & \mbox{05:21:59.59}  & \mbox{-48:16:06.00}  &      0.3000  & \mbox{05:22:14.81}  & \mbox{-48:18:00.73}  &        0.24  &            3.17  &        0.71  &        0.06  \\ 
\mbox{CXOUJ052215-481816}  & \mbox{05:22:15.70}  & \mbox{-48:18:18.00}  &      0.2960  & \mbox{05:22:14.81}  & \mbox{-48:18:00.73}  &        0.24  &            0.32  &        0.07  &        0.06  \\ i
\mbox{CXOUJ052246-481804}  & \mbox{05:22:46.61}  & \mbox{-48:18:03.99}  &      0.2100  & \mbox{05:22:47.78}  & \mbox{-48:18:05.77}  &        0.23  &            0.20  &        0.04  &       -0.02  \\
\mbox{CXOUJ104937-041728}  & \mbox{10:49:37.90}  & \mbox{-04:17:29.00}  &      0.2670  & \mbox{10:49:37.94}  & \mbox{-04:17:30.12}  &        0.22  &            0.02  &        0.00  &        0.05  \\
\mbox{CXOUJ105510-050414}  & \mbox{10:55:10.11}  & \mbox{-05:04:14.00}  &      0.6800  & \mbox{10:55:08.09}  & \mbox{-05:06:15.12}  &        0.60  &            2.08  &        0.83  &        0.08  \\ 
\mbox{CXOUJ105535-045930}  & \mbox{10:55:35.50}  & \mbox{-04:59:27.00}  &      0.6090  & \mbox{10:55:35.64}  & \mbox{-04:59:41.28}  &        0.54  &            0.24  &        0.09  &        0.07  \\
\mbox{CXOUJ140159-102301}  & \mbox{14:01:59.69}  & \mbox{-10:23:02.00}  &      0.4270  & \mbox{14:02:00.19}  & \mbox{-10:23:45.96}  &        0.42  &            0.74  &        0.25  &        0.00  \\
\hline
         \end{array}
      \]
\label{tab:knownclustersX}
   \end{table*}

\subsection{Comparison with weak lensing detections}

\cite{wittman06} detected eight massive clusters in 8.5 deg$^2$ of the DLS by using a shear-selection in WL maps. Of those eight clusters, they confirmed seven spectroscopically. In this work, we detect  all seven of these clusters with a redshift difference of $<0.08(1+z)$. In Table \ref{tab:knownclusters}, we list the spectroscopic redshift and our optical redshift estimate for comparison. 

\begin{table*}
      \caption{Clusters detected with WL and their optically detected counterparts}
      \[
\scriptsize
         \begin{array}{lllllllllllllrll}
            \hline\noalign{\smallskip}
\multicolumn{1}{c}{\rm Name}&
\multicolumn{1}{c}{\alpha (2000)}&
\multicolumn{1}{c}{\delta (2000)}&
\multicolumn{1}{c}{\rm z_{W06}}&
\multicolumn{1}{c}{\alpha (2000)_{A13}}&
\multicolumn{1}{c}{\delta (2000)_{A13}}&
\multicolumn{1}{c}{\rm z_{A13}}&
\multicolumn{1}{c}{\rm Offset}&
\multicolumn{1}{c}{\rm Offset}&
\multicolumn{1}{c}{\rm z_{W}-z_{A}}&
\multicolumn{1}{c}{\rm M_{A09}}&
\multicolumn{1}{c}{\rm M_{A13}}\\
\multicolumn{1}{c}{\rm }&
\multicolumn{1}{c}{}&
\multicolumn{1}{c}{}&
\multicolumn{1}{c}{}&
\multicolumn{1}{c}{}&
\multicolumn{1}{c}{}&
\multicolumn{1}{c}{}&
\multicolumn{1}{c}{'}&
\multicolumn{1}{c}{Mpc}&
\multicolumn{1}{c}{}&
\multicolumn{1}{c}{10^{14}M_{\odot}}&
\multicolumn{1}{c}{10^{14}M_{\odot}}\\
\hline\noalign{\smallskip}
\mbox{DLS0920.1+3029}  & \mbox{09:20:07.99}  & \mbox{+30:29:53.00}  &      0.2980  & \mbox{09:20:22.27}  & \mbox{+30:29:43.44}  &        0.25  &            3.08  &        0.72  &        0.05  &       8.45^{1.83}_{1.83} &        9.26^{2.03}_{2.03}  \\
\mbox{DLS0522.2-4820}  & \mbox{05:22:16.99}  & \mbox{-48:20:09.99}  &      0.2960  & \mbox{05:22:19.94}  & \mbox{-48:21:02.88}  &        0.29  &            1.01  &        0.26  &        0.01  &        3.94^{1.69}_{1.26} &      3.67^{1.64}_{1.71}  \\
\mbox{DLS1049.6-0417}  & \mbox{10:49:40.99}  & \mbox{-04:17:44.00}  &      0.2670  & \mbox{10:49:37.94}  & \mbox{-04:17:30.12}  &        0.22  &            0.79  &        0.17  &        0.05  &         0.70^{0.42}_{0.42} &     4.15^{2.17}_{2.23}  \\
\mbox{DLS1054.1-0549}  & \mbox{10:54:07.99}  & \mbox{-05:49:44.00}  &      0.1900  & \mbox{10:54:14.79}  & \mbox{-05:48:48.96}  &        0.21  &            1.92  &        0.36  &       -0.02  &           0.42^{0.28}_{0.28} &   3.46^{1.40}_{1.45}  \\
\mbox{DLS1402.0-1019}  & \mbox{14:02:03.00}  & \mbox{-10:19:44.00}  &      0.4270  & \mbox{14:02:00.19}  & \mbox{-10:23:45.96}  &        0.42  &            4.09  &        1.36  &        0.00  &      0.42^{0.42}_{0.42} &        4.11^{2.13}_{2.20}  \\
\mbox{DLS0916.0+2931}  & \mbox{09:15:60.00}  & \mbox{+29:31:34.00}  &      0.5300  & \mbox{09:15:56.97}  & \mbox{+29:33:47.52}  &        0.50  &            2.32  &        0.85  &        0.03  &     5.35^{3.24}_{2.82} &       3.61^{1.57}_{1.63}  \\
\mbox{DLS1055.2-0503}  & \mbox{10:55:12.00}  & \mbox{-05:03:43.00}  &      0.6800  & \mbox{10:55:03.77}  & \mbox{-05:02:09.24}  &        0.63  &            2.58  &        1.05  &        0.05  &      4.65^{1.97}_{1.97} &       4.66^{2.18}_{2.29}  \\
\hline
         \end{array}
      \]
\label{tab:knownclusters}
   \end{table*}   

\begin{figure}
\centering
\includegraphics[clip,angle=0,width=1.0\hsize]{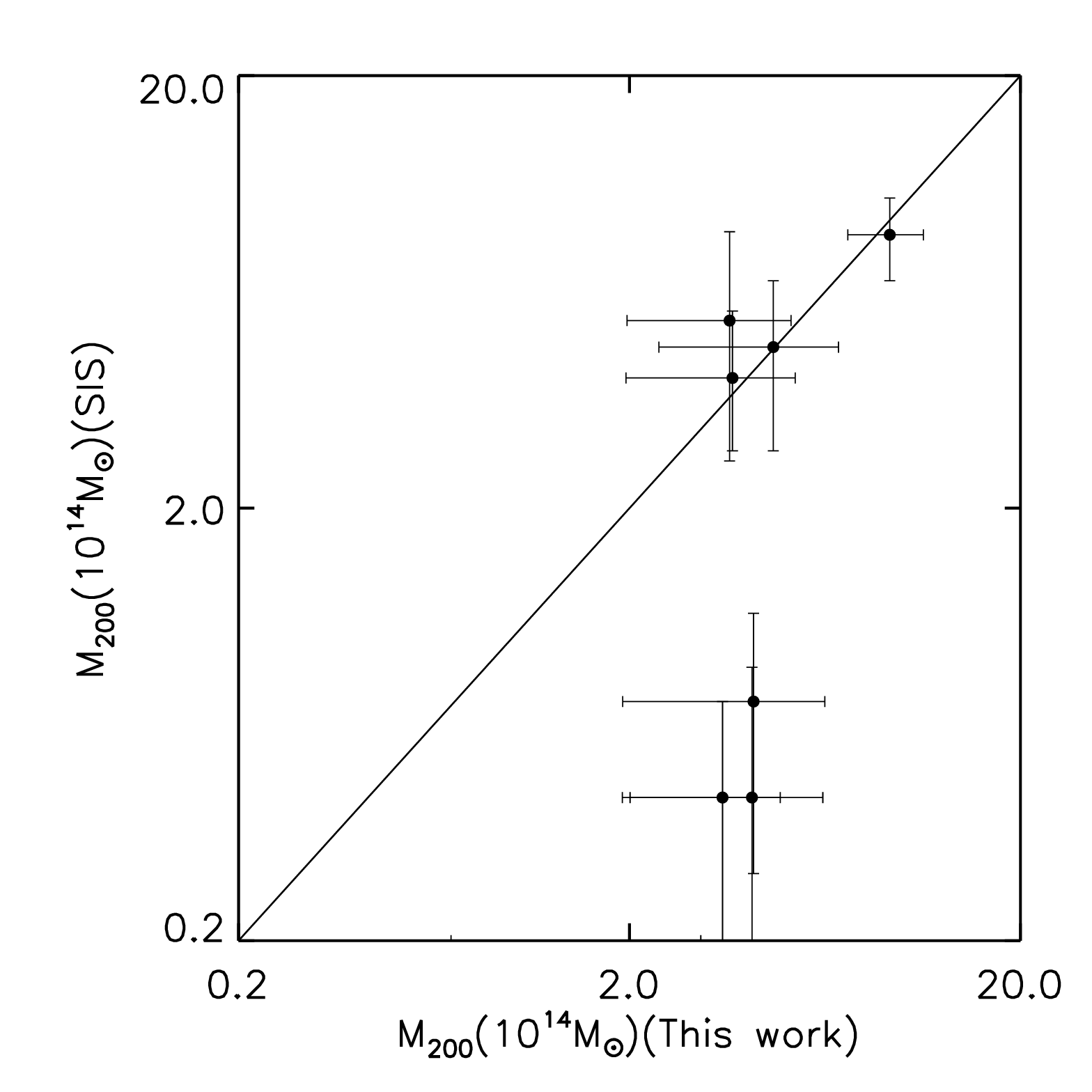} 
\caption{Comparison between WL masses obtained from Abate et al. 2009  and the optical masses obtained in this work.  Substructure is a  complicating factor in the clusters for which  the optical mass substantially exceeds the WL mass (see discussion  in text).}
\label{fig:massopticalAbate}
\end{figure}

We also compared the optical mass estimated for the DLS systems with the masses obtained from \cite{abate09}. They obtained masses for the seven candidates that \cite{wittman06} confirmed spectroscopically. In Figure \ref{fig:massopticalAbate}, we show the comparison between both masses. The A09 masses are the sum of the masses of all the  subclumps in each cluster. The mass values are listed in Table \ref{tab:knownclusters}. We find a good agreement between the estimates for the more massive clusters according to \cite{abate09}. The three least massive clusters in \cite{abate09} systematically have a larger mass estimate in this work. This could be caused by substructure. For instance, DLS1049.6-0417 has a  closer ($<$0.6 Mpc) group at redshift $\sim$ 0.13, DLS1054.1-0549 has a small group of galaxies at $\sim$ 0.11-0.16 at less than  $\sim$0.5 Mpc and DLS1402.0-1019 has a small group of galaxies at  z$\sim$ 0.5 separated by less than 0.9 Mpc, resulting in an overestimate of the mass in each of these clusters.

\subsection{Comparison with spectroscopic detections}

\cite{geller10} provide a list of spectroscopically detected groups and clusters from the Smithsonian Hectospec Lensing Survey (SHeLS, \citealt{geller05}) which overlaps DLS field F2.  With a spectroscopic survey complete to $R=20.3$, they were able to find clusters up to $z\sim 0.55$ and groups at lower redshift.  In Table  \ref{tab:knownclustersSHELS}, we list our counterparts to their detections.  We detect  90\% of their detections.  Section \S\ref{clusterprops} indicates that we are detecting down to $\sim 2.4 \times 10^{14} M_\odot$ at the redshifts SHeLS probes, and the SHeLS cross-matching results are consistent with this. In fact, the two clusters that we do not find are detected initially but excluded after the $\Lambda_{CL} \ge 40$ cut (see Section \S\ref{DLSopt}). In comparing the efficacy of photometric and spectroscopic cluster selection there are two competing effects. The photometry goes much deeper and contains many more galaxies than the spectroscopy ($\sim$1 million vs. $\sim$10,000 in this field), but the limited redshift resolution provided by photometric redshifts allows clusters to be smeared out in redshift space, and this may allow poorer clusters to escape our detection threshold.  Therefore, our sample goes further down the mass function in general (see below) but individual clusters can scatter out of our sample.

\begin{figure}
\centering
\includegraphics[clip,angle=0,width=1.0\hsize]{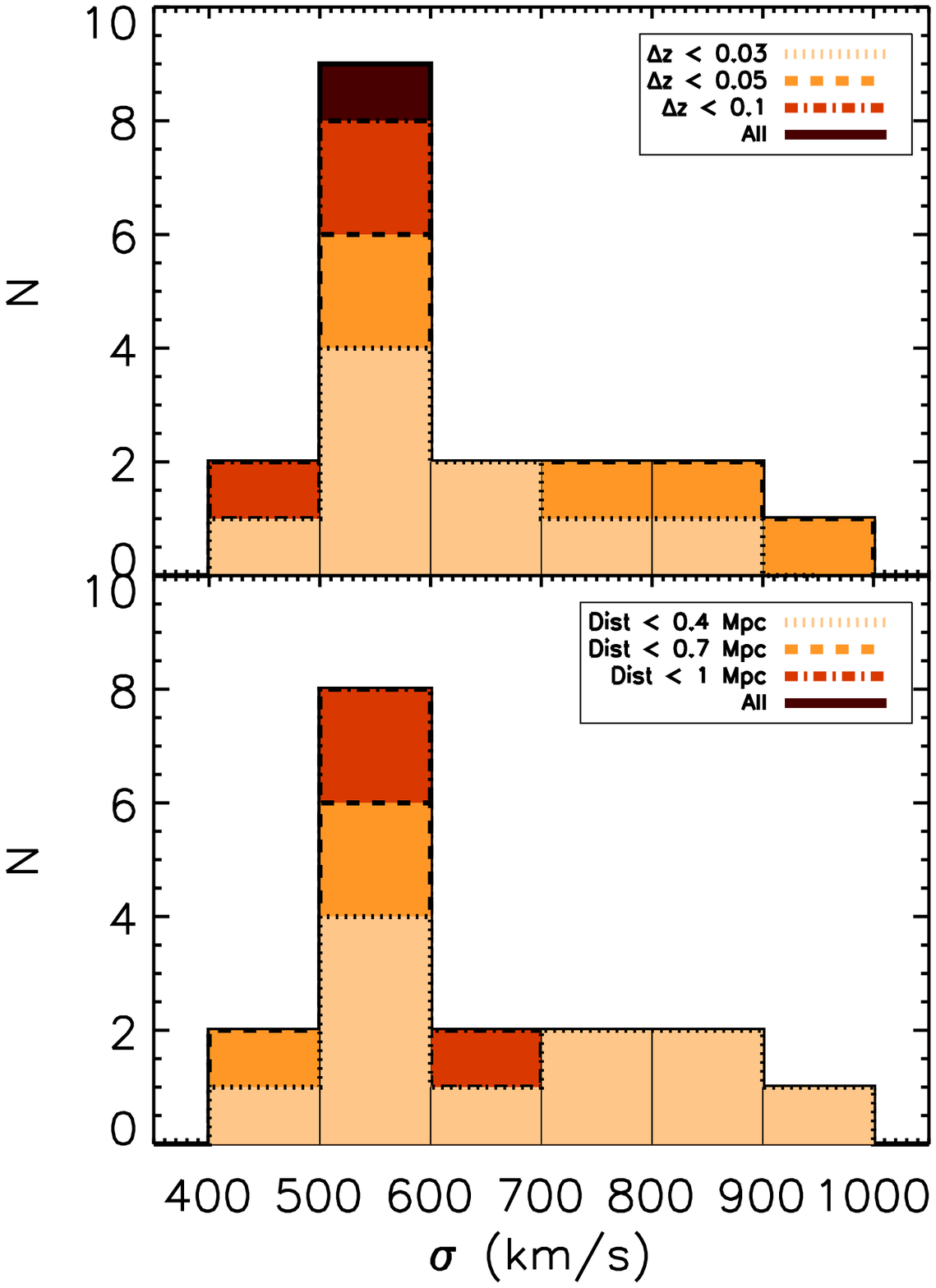} 
\caption{Distribution of  $\sigma_3$ for the clusters in Geller et al. 2010 that we detect in this work. Different lines  and colors indicate different redshift tolerances for fixed physical separation of 1 Mpc (upper plot) and different physical separation for a fixed redshift tolerance of 0.1 (bottom plot).}
\label{fig:histosigma}
\end{figure}

In Figure \ref{fig:histosigma}, we show the distribution of SHeLS $\sigma_3$, the rest-frame velocity dispersion within 3 arc minutes, for SHeLS clusters that we detect, for various matching criteria. In the top panel, we fix the tolerance in projected physical separation at  1 Mpc, and vary the redshift tolerance.  In the bottom plot, we vary the physical separation while maintaining a redshift tolerance of 0.1.  We can see very clearly that the detections with larger offsets both in redshift and position are the ones with smaller velocity dispersion values.   Clusters with larger velocity dispersions show an excellent agreement with the SHeLS detections.

\begin{figure}
\centering
\includegraphics[clip,angle=0,width=1.0\hsize]{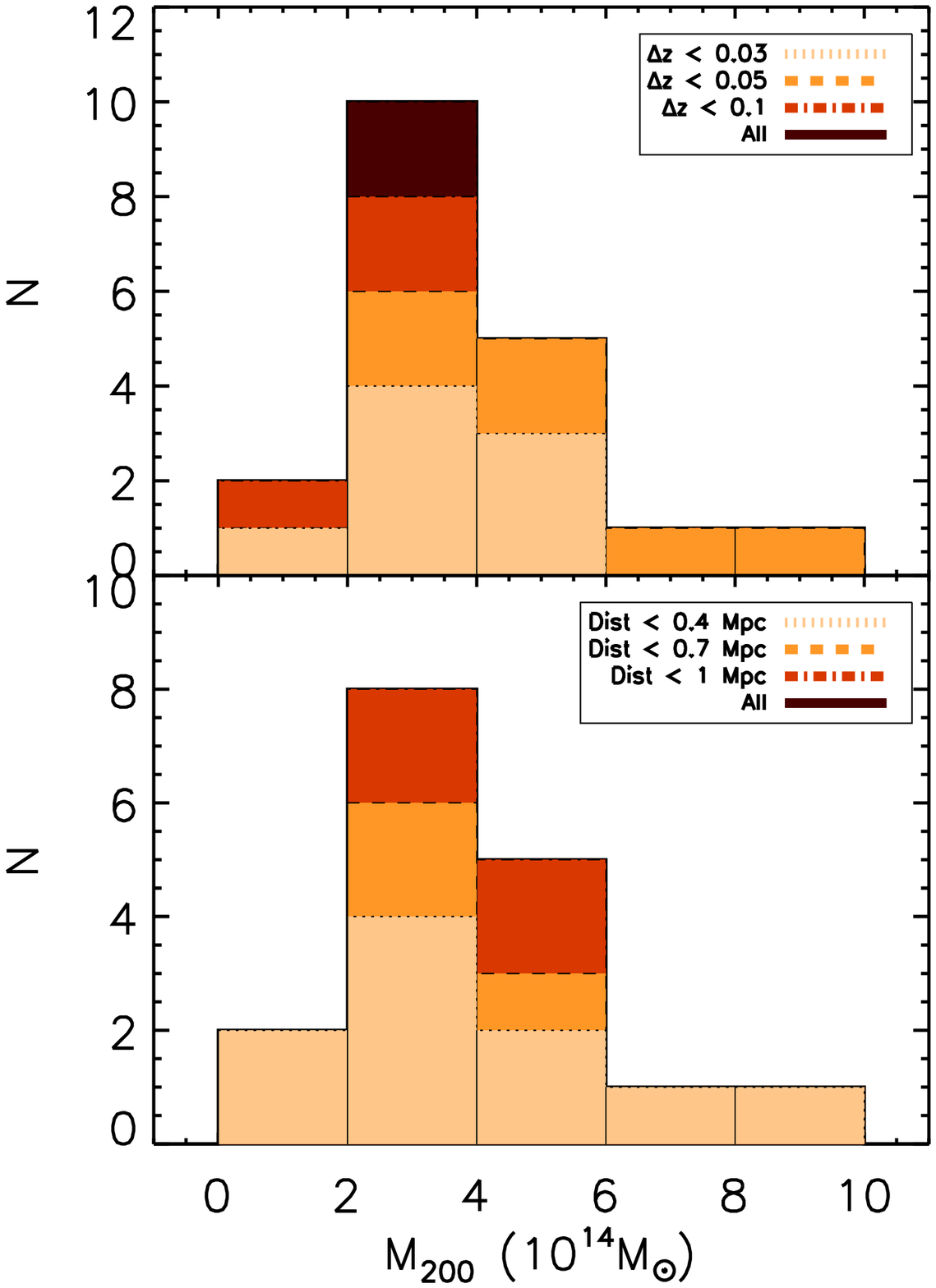} 
\caption{Distribution of  $M_{200}$ for the clusters in this work that Geller et al. 2010 detected for different redshift tolerances (upper plot) and physical separations (bottom plot). The lines and colors are the same as in Figure \ref{fig:histosigma}.}
\label{fig:histomassG}
\end{figure}

Similarly, we also examined the \cite{geller10} counterparts to our detections at  $z<0.55$ (as explained in A12, this may differ slightly from our counterparts to the \cite{geller10} detections). In Figure \ref{fig:histomassG}, we show the distribution of these clusters as a function of our $M_{200}$.  As before, the top panel shows the effect of varying the redshift matching tolerance and the bottom panel shows the effect of varying the positional matching tolerance. As before, we observe that clusters with larger redshift and position offsets tend to have lower mass. On the contrary, the higher mass clusters are found to have a smaller redshift and distance tolerances.

\begin{figure}
\centering
\includegraphics[clip,angle=0,width=1.0\hsize]{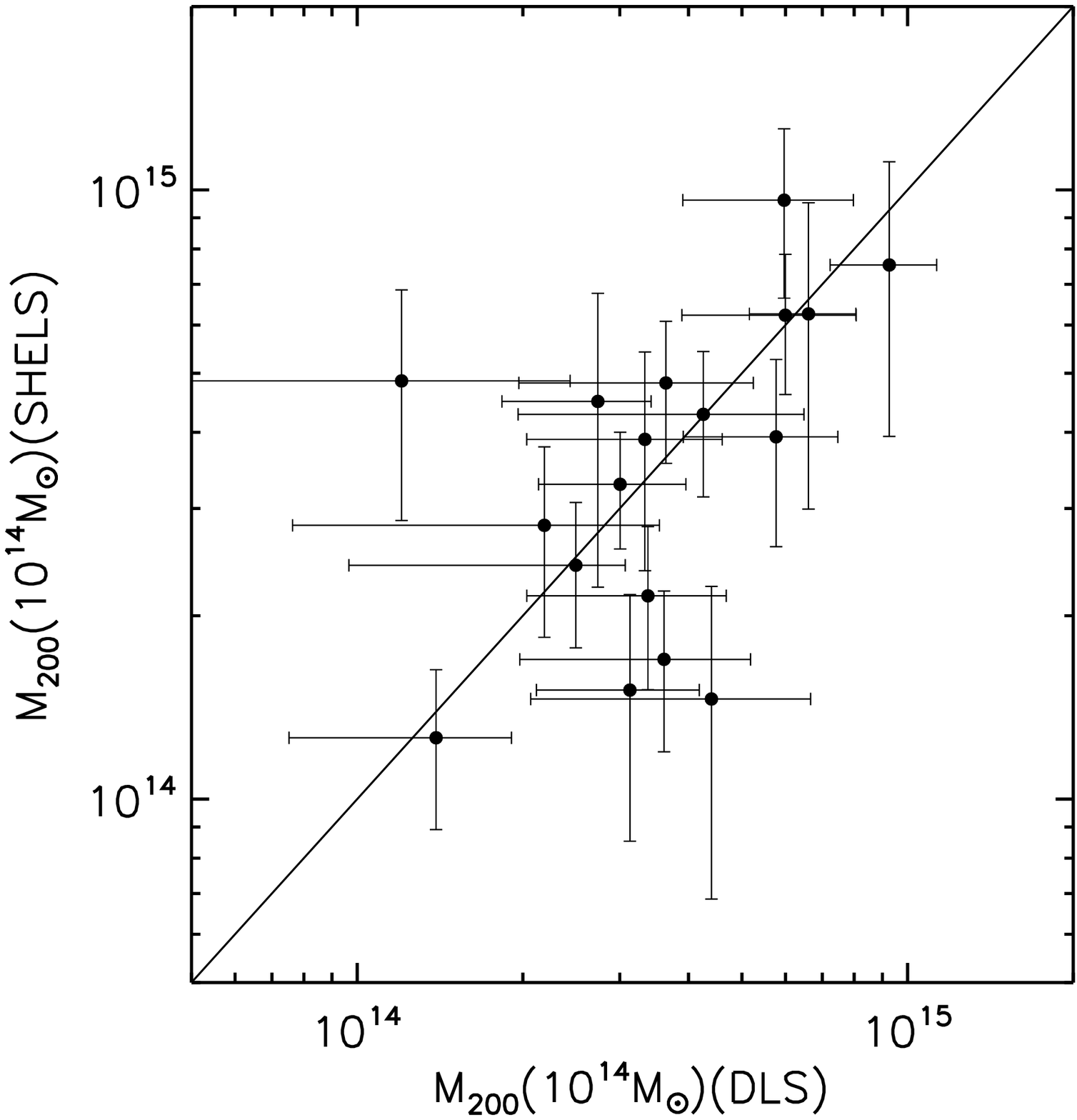} 
\caption{Comparison of the masses inferred from the optical richness  in this work with the dynamical masses from Geller et al. 2010. The solid line illustrates the one-to-one mass relationship. The  scatter of 0.24 dex is consistent with the nominal uncertainties  in each catalog and similar to the scatter found when comparing  richness-inferred masses to masses inferred from X-ray or  weak-lensing \citep{andreon10,wen12}.}
\label{fig:massG}
\end{figure}

Finally, Figure \ref{fig:massG} compares the mass estimates that we obtain in this work with the mass estimates obtained from the velocity dispersion $\sigma_3$ given by \cite{geller10} by applying the virial theorem assuming that these clusters are spherical \citep{girardi98}. The sample agrees well within the errors when they are matched to the SHeLS sample or when we match the SHeLS sample to the DLS,  obtaining a dispersion of 0.24 dex for both samples. Other works  \citep{andreon10,wen12}, found a similar dispersion when comparing optical richness to $M_{200}$ estimated from X-ray or weak-lensing.

\begin{table*}
      \caption{Clusters detected in the SHELS and their optically detected counterparts}
      \[
\scriptsize
         \begin{array}{llllllllllll}
            \hline\noalign{\smallskip}
\multicolumn{1}{c}{\rm Name}&
\multicolumn{1}{c}{\alpha (2000)}&
\multicolumn{1}{c}{\delta (2000)}&
\multicolumn{1}{c}{\rm z_{S}}&
\multicolumn{1}{c}{\alpha (2000)_{A13}}&
\multicolumn{1}{c}{\delta (2000)_{A13}}&
\multicolumn{1}{c}{\rm z_{A13}}&
\multicolumn{1}{c}{\rm Offset}&
\multicolumn{1}{c}{\rm Offset}&
\multicolumn{1}{c}{\rm z_{S}-z_{A}}&
\multicolumn{1}{c}{\rm M_{S}}&
\multicolumn{1}{c}{\rm M_{A13}}\\
\multicolumn{1}{c}{}&
\multicolumn{1}{c}{}&
\multicolumn{1}{c}{}&
\multicolumn{1}{c}{}&
\multicolumn{1}{c}{}&
\multicolumn{1}{c}{}&
\multicolumn{1}{c}{}&
\multicolumn{1}{c}{'}&
\multicolumn{1}{c}{Mpc}&
\multicolumn{1}{c}{}&
\multicolumn{1}{c}{10^{14}M_{\odot}}&
\multicolumn{1}{c}{10^{14}M_{\odot}}\\
\hline\noalign{\smallskip}
\mbox{J0915.1+2954}  & \mbox{09:15:03.51}  & \mbox{+29:54:09.00}  &      0.1319  & \mbox{09:15:05.73}  & \mbox{+29:56:21.84}  &        0.17  &            2.27  &        0.32  &       -0.03  &       1.46 \pm 0.77  &            4.40^{2.26}_{2.34}  \\
\mbox{J0916.0+3028}  & \mbox{09:15:57.09}  & \mbox{+29:49:42.01}  &      0.1844  & \mbox{09:15:49.46}  & \mbox{+29:50:02.40}  &        0.21  &            1.69  &        0.31  &       -0.03  &       1.51 \pm 0.66  &            3.13^{1.05}_{1.01}  \\
\mbox{J0916.2+2949}  & \mbox{09:16:10.90}  & \mbox{+29:48:44.00}  &      0.5343  & \mbox{09:16:14.88}  & \mbox{+29:49:41.52}  &        0.52  &            1.29  &        0.48  &        0.01  &       9.62 \pm 2.98  &            5.97^{2.01}_{2.06}  \\
\mbox{J0916.3+2916}  & \mbox{09:16:19.20}  & \mbox{+29:15:47.00}  &      0.5347  & \mbox{09:16:18.31}  & \mbox{+29:17:40.92}  &        0.52  &            1.91  &        0.71  &        0.02  &       4.28 \pm 1.15  &            4.26^{2.23}_{2.30}  \\
\mbox{J0916.7+2920}  & \mbox{09:16:40.10}  & \mbox{+29:19:52.00}  &      0.2158  & \mbox{09:16:37.25}  & \mbox{+29:21:06.12}  &        0.21  &            1.38  &        0.29  &        0.00  &       2.42 \pm 0.65  &            2.50^{0.57}_{1.53}  \\
\mbox{J0916.8+2908}  & \mbox{09:16:49.99}  & \mbox{+29:08:19.00}  &      0.3356  & \mbox{09:16:43.42}  & \mbox{+29:08:15.00}  &        0.53  &            1.44  &        0.41  &       -0.19  &       3.29 \pm 0.71  &            3.00^{0.95}_{0.87}  \\
\mbox{J0916.9+3003}  & \mbox{09:16:56.71}  & \mbox{+30:03:08.00}  &      0.3189  & \mbox{09:17:03.26}  & \mbox{+30:01:20.64}  &        0.24  &            2.28  &        0.51  &        0.08  &       2.16 \pm 0.64  &            3.38^{1.31}_{1.34}  \\
\mbox{J0918.1+3038}  & \mbox{09:18:05.81}  & \mbox{+30:37:48.00}  &      0.3970  &     -    &  -    &    -    &     -      &    -    &    -   &            4.93      \pm     2.34  &    -   \\
\mbox{J0918.2+3057}  & \mbox{09:18:09.79}  & \mbox{+30:56:56.00}  &      0.4244  & \mbox{09:18:10.15}  & \mbox{+30:56:06.00}  &        0.43  &            0.84  &        0.28  &       -0.00  &       4.86 \pm 1.99  &            1.20^{1.23}_{1.01}  \\
\mbox{J0918.3+3024}  & \mbox{09:18:16.01}  & \mbox{+30:24:07.00}  &      0.1241  & \mbox{09:18:07.15}  & \mbox{+30:23:03.84}  &        0.21  &            2.18  &        0.29  &       -0.09  &       1.26 \pm 0.37  &            1.39^{0.52}_{0.64}  \\
\mbox{J0918.6+2953}  & \mbox{09:18:38.59}  & \mbox{+29:53:22.00}  &      0.3178  & \mbox{09:18:34.34}  & \mbox{+29:53:18.60}  &        0.27  &            0.92  &        0.23  &        0.05  &       4.82 \pm 1.26  &            3.64^{1.61}_{1.67}  \\
\mbox{J0919.6+3032}  & \mbox{09:19:33.29}  & \mbox{+30:31:59.00}  &      0.4273  & \mbox{09:19:34.27}  & \mbox{+30:30:28.44}  &        0.49  &            1.52  &        0.51  &       -0.06  &       3.89 \pm 1.52  &            3.33^{1.27}_{1.30}  \\
\mbox{J0920.1+3010}  & \mbox{09:20:03.60}  & \mbox{+30:10:06.00}  &      0.4263  & \mbox{09:19:55.51}  & \mbox{+30:11:28.32}  &        0.46  &            2.22  &        0.74  &       -0.04  &       2.81 \pm 0.97  &            2.19^{1.35}_{1.42}  \\
\mbox{J0920.4+3030*}  & \mbox{09:20:22.51}  & \mbox{+30:30:29.00}  &      0.3004  & \mbox{09:20:22.27}  & \mbox{+30:29:43.44}  &        0.25  &            0.76  &        0.18  &        0.05  &       7.53 \pm 3.59  &            9.26^{2.03}_{2.03}  \\
\mbox{J0920.9+3029*}  & \mbox{09:20:55.61}  & \mbox{+30:28:38.00}  &      0.2915  & \mbox{09:20:52.89}  & \mbox{+30:28:46.92}  &        0.25  &            0.60  &        0.14  &        0.04  &       6.26 \pm 3.27  &            6.61^{1.45}_{1.45}  \\
\mbox{J0921.0+2942}  & \mbox{09:20:59.59}  & \mbox{+29:42:00.00}  &      0.2964   &     -    &  -    &    -    &     -      &    -    &    -   &           1.85       \pm    0.65    &    -   \\
\mbox{J0921.2+3028}  & \mbox{09:21:12.70}  & \mbox{+30:28:08.00}  &      0.4265  & \mbox{09:21:11.18}  & \mbox{+30:27:45.00}  &        0.41  &            0.50  &        0.16  &        0.02  &       6.23 \pm 1.61  &            6.00^{2.07}_{2.10}  \\
\mbox{J0921.3+2946}  & \mbox{09:21:13.90}  & \mbox{+29:45:37.00}  &      0.3834  & \mbox{09:21:24.70}  & \mbox{+29:46:10.92}  &        0.38  &            2.41  &        0.74  &        0.01  &       4.50 \pm 2.27  &            2.74^{0.68}_{0.90}  \\
\mbox{J0921.4+2958}  & \mbox{09:21:24.91}  & \mbox{+29:58:12.00}  &      0.4318  & \mbox{09:21:17.59}  & \mbox{+29:59:14.64}  &        0.42  &            1.90  &        0.63  &        0.01  &       3.93 \pm 1.34  &            5.77^{1.70}_{1.86}  \\
\mbox{J0923.6+2929}  & \mbox{09:23:37.99}  & \mbox{+29:28:35.00}  &      0.2216  & \mbox{09:23:36.91}  & \mbox{+29:29:01.32}  &        0.21  &            0.50  &        0.10  &        0.01  &       1.70 \pm 0.50  &            3.61^{1.57}_{1.64}  \\
\hline
         \end{array}
      \]
\begin{flushleft} 
$^*$: The velocity dispersion of these two clusters are estimated from Table 2 in \cite{geller10}
\end{flushleft}
\label{tab:knownclustersSHELS}
   \end{table*}

\subsection{Comparison with weak lensing maps}

For each field we created matched pairs of optical detection and weak lensing maps.  For the weak lensing maps we attempted to replicate the sliced redshift search of the BCF as follows.  For a series of putative lens redshifts (0.2, 0.3, 0.4, 0.5, 0.6 and 0.7), we made convergence maps which were optimized for each lens redshift by weighting the sources according to their distance ratio assuming a lens at that redshift.  However, unlike the optical search, each lens will appear in each convergence map because the distance-ratio weights change quite slowly with lens redshift.  We attempt to remove this effect by converting the convergence maps into signal-to-noise maps and, at each location in sky coordinates, taking the maximum along the
redshift axis.  The result is not a standard lensing map, but it does replicate the basic features of the BCF.  Each cluster should appear on this map at its sky location and with pixel values proportional to its mass and to the effective distance ratio of DLS source galaxies optimized for that lens redshift \citep{dawson12}.

\begin{figure}
\centering
\includegraphics[clip,angle=0,width=1.0\hsize]{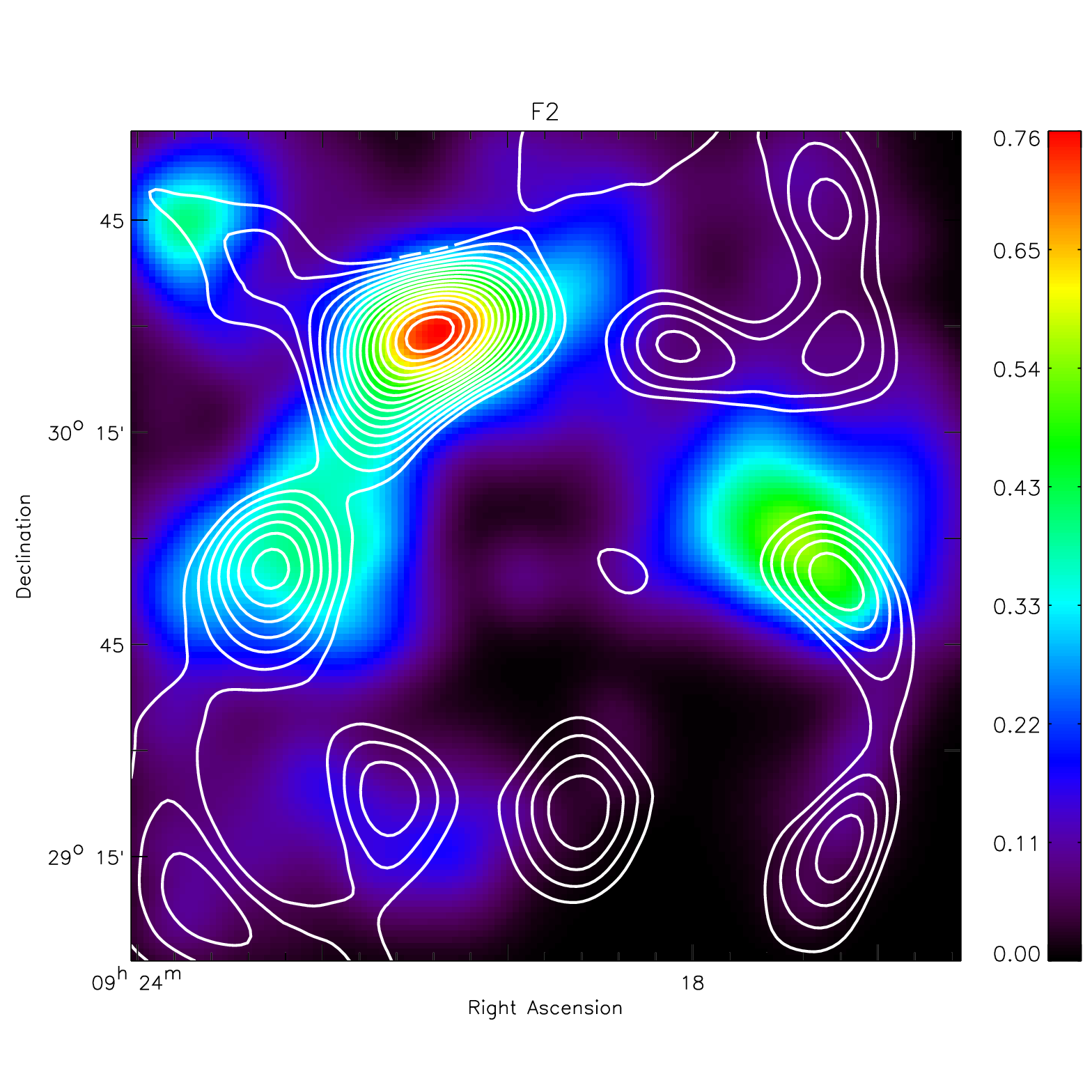} 
\caption{Synthetic weighted surface mass density map for the cluster detections in  DLS field F2 in linear scale (colour map) in arbitrary units. The white contours correspond to the smoothed WL signal-to-noise map with linear contour intervals spaced from 0 to 5. Regions without contours are underdense according to the lensing analysis. The other four DLS fields can be found in the Appendix.}
\label{fig:DLSWLcontF2}
\end{figure}

We create a representation of what the optical cluster detections should look like on this map as follows.  We represent each cluster by a two-dimensional Gaussian centered at the cluster's sky coordinates and with a 1$\sigma$ width of 1.5 Mpc at the redshift of the cluster. We normalize the Gaussian to the cluster's mass as inferred from its richness via the \cite{dong08} relation, times the effective distance ratio of source galaxies used for that lens redshift in the weak lensing analysis described above. In Figure  \ref{fig:DLSWLcontF2}, the background color map represents the distance-weighted mass map synthesized from the optical detections in Field F2, and the white contours represent the smoothed WL signal-to-noise map, with underdense regions omitted. Figures  \ref{fig:DLSWLcontF1},  \ref{fig:DLSWLcontF3},  \ref{fig:DLSWLcontF4} and  \ref{fig:DLSWLcontF5} in the Appendix show the corresponding maps for the DLS F1, F3, F4 and F5 fields respectively. The weak lensing maps are noisy; one indication of this is that the most significant peak is detected at only about $5\sigma$. Therefore, many of the higher peaks should match up while many of the lower peaks should not, and this expectation is borne out.

\begin{figure}
\centering
\includegraphics[clip,angle=0,width=1.0\hsize]{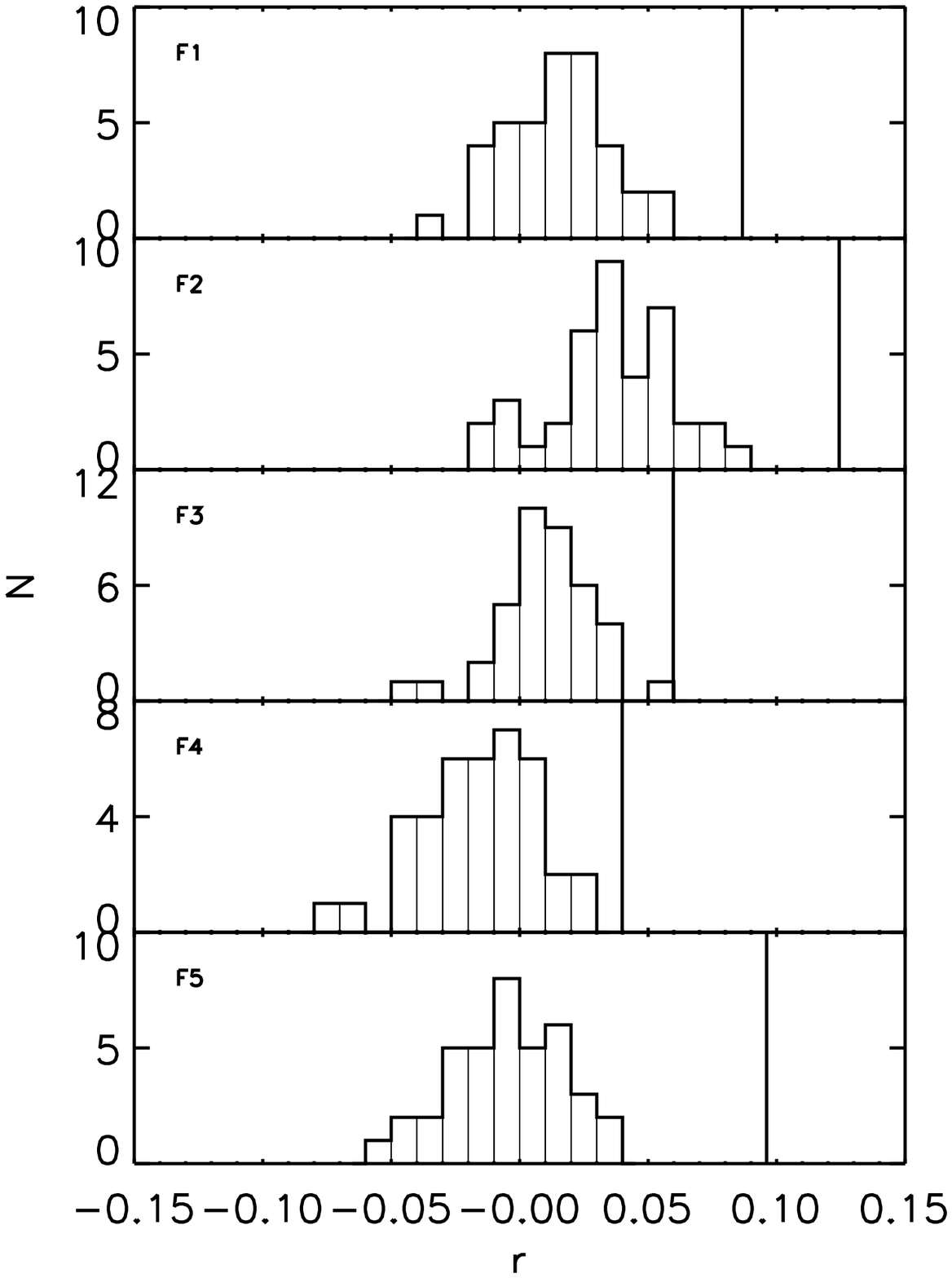} 
\caption{Distribution of the correlation coefficient of the null-hypothesis of correlation between optically detected cluster maps and weak-lensing maps for a control sample of unrelated weak lensing maps. In each field, the vertical line indicates the optical-WL correlation coefficient for the true WL map for each different field.}
\label{fig:correlationprob}
\end{figure}

We quantified the level of correlation, finding that the correlation coefficient of the optical and weak lensing maps ranges from 0.04 to 0.12, depending on the field, with a median of 0.09.  In other words, about 10\% of the variance in the weak lensing maps is explained by variance in the optical map, consistent with the statement that the weak lensing maps are dominated by noise.  The observed correlation coefficients, although low, are real.  We confirmed this by correlating the optical map in each field with a control sample of unrelated weak lensing maps---rotations and transpositions of the weak lensing maps, including those from other fields.  The distributions of these correlation coefficients clearly exclude the true correlation coefficients, as shown in Fig. \ref{fig:correlationprob}.  Therefore, we have high confidence that a fraction of the variance in the weak lensing maps is explained by the optical detections, even though that fraction is small due to noise in the weak lensing maps.

\section{Cluster sample redshift and mass distributions}
\label{clusterprops}

Many works in the literature have been devoted to constraining cosmological parameters with cluster counts \citep{henry00,mantz08,mantz10,rozo10} using large cluster samples. In this work, we have detected clusters in a 20 $deg^2$ sampling a wide range in mass and redshift. While we do not attempt to set cosmological constraints due to the sample variance, as a further check on the integrity of our sample we compare the recovered mass function with that predicted by LCDM.

\begin{figure}
\centering
\includegraphics[clip,angle=0,width=1.0\hsize]{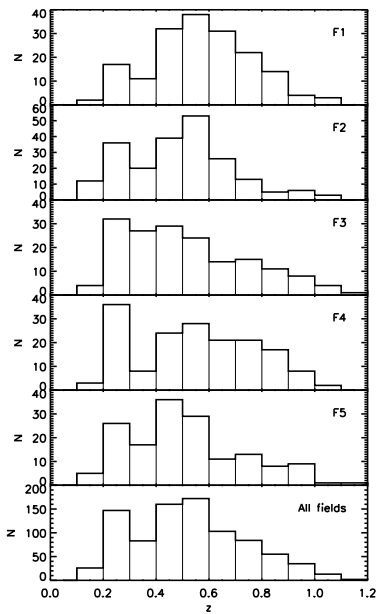} 
\caption{Differential redshift distribution for the clusters found in each field of the DLS.}
\label{fig:histoz}
\end{figure}

In Figure  \ref{fig:histoz}, we show the differential redshift distribution of the clusters detected in each DLS field.  Across the entire survey, this distribution peaks at $0.5<z<0.6$, but there is substantial variation from field to field due to sample variance.  For example, F2 appears to be quite rich even in a simple visual inspection of its images, and this is reflected in the different vertical scale in the F2 panel. Moreover, \cite{geller10}  found multiple clusters in their spectroscopic survey in each of the richest F2 bins and \cite{jee13} also confirmed the richness of F2 with cosmic shear statistics.

\begin{figure}
\centering
\includegraphics[clip,angle=0,width=1.0\hsize]{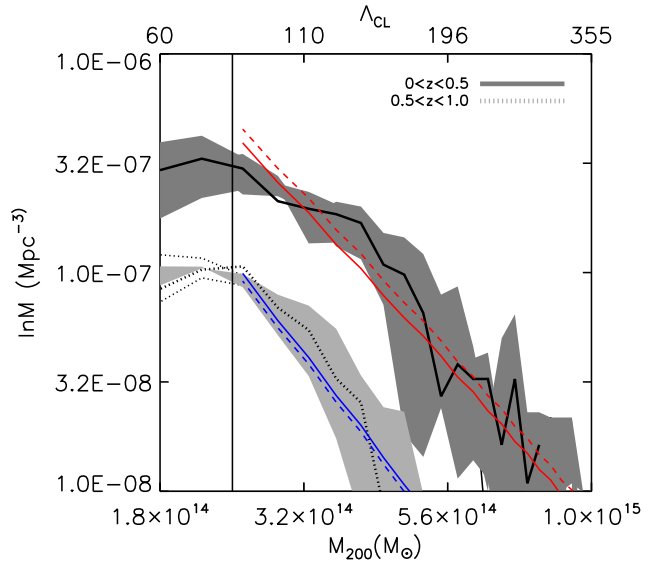} 
\caption{Differential distribution of cluster mass in the DLS for different redshift bins ($z\le 0.5$, solid line and $0.5<z\le 1$, dotted line). The shaded areas show the uncertainties in converting richness to mass according to Dong et al. 2008 and the sample variance. For comparison, CDM predictions from simulations by Jenkins et al. 2001 (solid color lines) and Tinker et al. 2008 (dashed color lines) are overplotted with solid lines for the same redshift bins as the data.}
\label{fig:histoMassz}
\end{figure}

To examine the mass function and its redshift evolution, we first correct for the different volume probed at different redshifts. Figure  \ref{fig:histoMassz} shows the differential number of clusters per cubic Mpc as a function of mass in two redshift bins:  0-0.5 and 0.5-1.0.  The evolution with redshift is very striking, as many more high-mass clusters are detected (per unit volume probed) at low redshifts.  This cannot be a selection effect, because selection effects would have the opposite sign, favoring massive clusters over low-mass clusters at high redshift.  We compared the mass function with numerical predictions. The solid color lines are the \cite{jenkins01} model whereas the dashed color lines refer to the \cite{tinker08} predictions for $\Delta$ =200. The results agree with both models between $2.4\times 10^{14}$ and $1\times 10^{15} M_{\odot}$ while they seem to overestimate the number of clusters with respect to the models for masses lower than this limit. Hence, we define the completeness limit as the mass down to which the measurements are consistent with the theoretical models (vertical line), this limit being 2.4$\times 10^{14}$ M$_\odot$.

The observed differential distribution of mass in clusters is consistent with theoretical predictions, which supports our claim  that the sample is highly complete in this mass and redshift range.

\section{BCG evolution in the DLS}

In the last few years, extensive studies on the evolution of the brightest cluster galaxies (BCGs)  have appeared in the  literature.  These objects are the largest and brightest galaxies in the universe and their formation is expected to be directly related to the formation of the cluster. More massive, larger, and more dominant BCGs tend to live in the most massive and luminous clusters. Indeed, recent work  \citep{ree07,whiley08,stott08,bildfell08,liu09,bernardi09,ruszkowski09,odea10,ascaso11,tonini12,lidman12,lidman13,ascaso13} reveals significant dependence on the cluster environment. However, the role that the cluster environment is playing in such galaxies is still a matter of debate.

We selected the BCG of each cluster by considering all the galaxies enclosed within 1.0 Mpc and within a photometric redshift range of $0.08(1+z_c)$  and photometric redshift odds $>$  0.8. Then, we fit the red sequence ($R-z$ and $B-R$ color versus $R$ magnitude) to early type galaxies, which we defined as having a best-fit BPZ type $t_B\le 2$ \citep{benitez00}. We applied a 3$\sigma$-clipping algorithm to the red sequence fit and finally, we selected the brightest of the galaxies within 3$\sigma$ of the cluster red sequence.  Two effects might make BCGs confused with field or  saturated galaxies. First, catastrophic photometric redshift errors  occasionally include obvious foreground galaxies in the selection of  the cluster population. Second, the saturation of bright galaxies in  the DLS invites confusion with bright stars, which morphologically differ from the PSF and may therefore contaminate the catalog.  For  these reasons, we perform a visual classification of our selection  sample. We define class A BCG candidates as those which clearly have  BCG morphology (i.e. early-type galaxies, large halos), or which are clearly the best BCG candidate in the region; class B  candidates as those which are plausible but which are not the only  plausible candidate in the cluster; and class C candidates are  clearly spurious (stars or foreground galaxies). We also included a saturation flag indicating whether the BCG was  saturated due to long exposures. 

Furthermore, another source of uncertainty needs to be taken into account. Some BCGs have been reported to have bluer colors than the main red sequence \citep{wen11,liu12,postman12b,ascaso13}. To take this  fact into account, we set a color flag indicating the existence of bluer and brighter galaxies than the BCG candidate for a particular cluster. We found 27.3\% of the whole sample have a bluer and brighter candidate. We also visually examined these blue candidates and classified them into one of the previous categories. Blue candidates classified as A were considered as the cluster BCG. The final classification for the whole sample was 48.53\%, 41.50\% and 9.98\% as A, B and C respectively.  

In the following analysis, we impose several cuts to reduce the possibility of bias.  At the high-redshift end, properties of clusters at $z>0.8$ become much more uncertain because most galaxies fainter galaxies than L* are not detectable.  At the low-redshift end, saturation affects many BCG candidates at z$\le$0.25, so we limit the following analysis to $0.25 < z \le 0.8$.  Within these limits, a few BCGs are still saturated, so we performed two-sided  Kolmogorov-Smirnov (KS) tests on each of the seven cluster  properties examined below to determine if excluding the saturated BCGs resulted in any significant differences.  We found no significant differences, so we impose only the $0.25 < z \le 0.8$  and not an additional saturation cut.  In addition, we used only  clusters with $M_{200} \ge 2.4 \times10^{14} M_{\odot}$ in order to  select a complete sample in mass and avoid biases (see Figure  \ref{fig:histoMassz}).  This yielded a subsample of 285 BCGs and their  host clusters of which 258 BCGs are classified as A+B (of these, 159 are not saturated). We further determined that there are no significant differences between the A+B and the A sample using the same suite of KS tests.  Therefore, the following analysis uses the 258 systems which are classified as A or B, which have  $M_{200} \ge  2.4\times10^{14} M_{\odot}$, and which lie in the redshift range $0.25 < z \le 0.8$.

We now examine the correlation of BCG properties with cluster properties. We chose the size of the BCG, estimated from the isophotal area, the absolute magnitude (k de-corrected and corrected for passive luminosity evolution, \citealt{blanton03}) and redshift of the BCG. As for the cluster, we selected the redshift, $M_{200}$, $R_{200}$ and the total luminosity, $L_R$.  In Figure \ref{fig:BCGhost}, we show the relation between BCG properties and host cluster properties.  We performed a Spearman test to investigate if cluster and BCGs parameters deviates from the null hypothesis of no correlation. In Table \ref{tab:corr}, we list the number of standard deviations by which the sum-squared difference of rank deviates from the null hypothesis expected value. For clarity, we have not included the error bars in the plot but they are included in Table \ref{tab:corr}.

\begin{table*}
      \caption{Significance of the Spearman Test for the structural parameters of the BCGs and their host cluster properties in the DLS sample.}
      \[
         \begin{array}{lccccc}
            \hline\noalign{\smallskip}
\multicolumn{1}{c}{\rm }&
\multicolumn{1}{c}{\rm z_{cluster} }&
\multicolumn{1}{c}{\rm R_{200}}&
\multicolumn{1}{c}{\rm M_{200}}&
\multicolumn{1}{c}{\rm Lum}\\
\hline\noalign{\smallskip}
\rm M_R       &    4.503 & 3.644 & 2.144 & 2.466   \\
\rm z       & 13.744 &    2.571 &  3.106 & 4.374   \\
\rm size  & 4.776 & 0.861 & 0.228 & 1.346  \\

         \end{array}
      \]
\label{tab:corr}
   \end{table*}

We consider a correlation to be significant if the absolute value of the significance of the Spearman test is larger than 3.  Table \ref{tab:corr} shows that, according to this criterion, the BCG $M_R$ is positively correlated with cluster redshift and $R_{200}$ and the BCG size is negatively correlated with cluster redshift in agreement with other works up to redshift $\sim 1$ (e.g. \citealt{bernardi09,liu09,ascaso11,ascaso13,lidman13}).  While it is well known that the most massive clusters host the brightest BCGs (see similar results for more massive clusters in A12 (up to z$\sim$ 0.6) and \cite{wen11} (up to z$\sim$1.6)), we see this tendency at a level which is not quite significant probably due to the large uncertainties in the mass measurements.

\begin{figure}
\centering
\includegraphics[clip,angle=0,width=1.0\hsize]{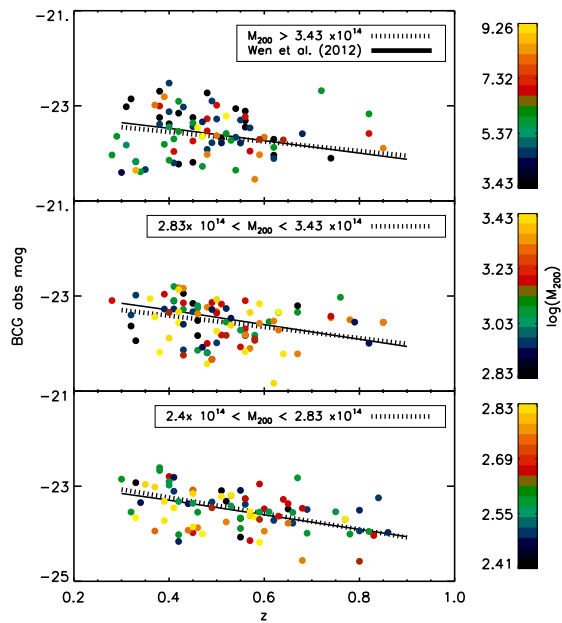} 
\caption{Absolute magnitude versus redshift for different masses bins:  $2.4\times 10^{14} M_{\odot}<M_{200}<2.83\times 10^{14} M_{\odot}$, $2.83\times 10^{14} M_{\odot}<M_{200}<3.43\times 10^{14} M_{\odot}$ and  $M_{200}>3.43\times 10^{14} M_{\odot}$ from top to bottom. Each panel is color-coded according to the host mass halo mass. The dotted and solid lines are our fits and the Wen et al. (2012) fits for each of the mass bins respectively.}
\label{fig:zmag}
\end{figure}

\begin{figure}
\centering
\includegraphics[clip,angle=0,width=0.75\hsize]{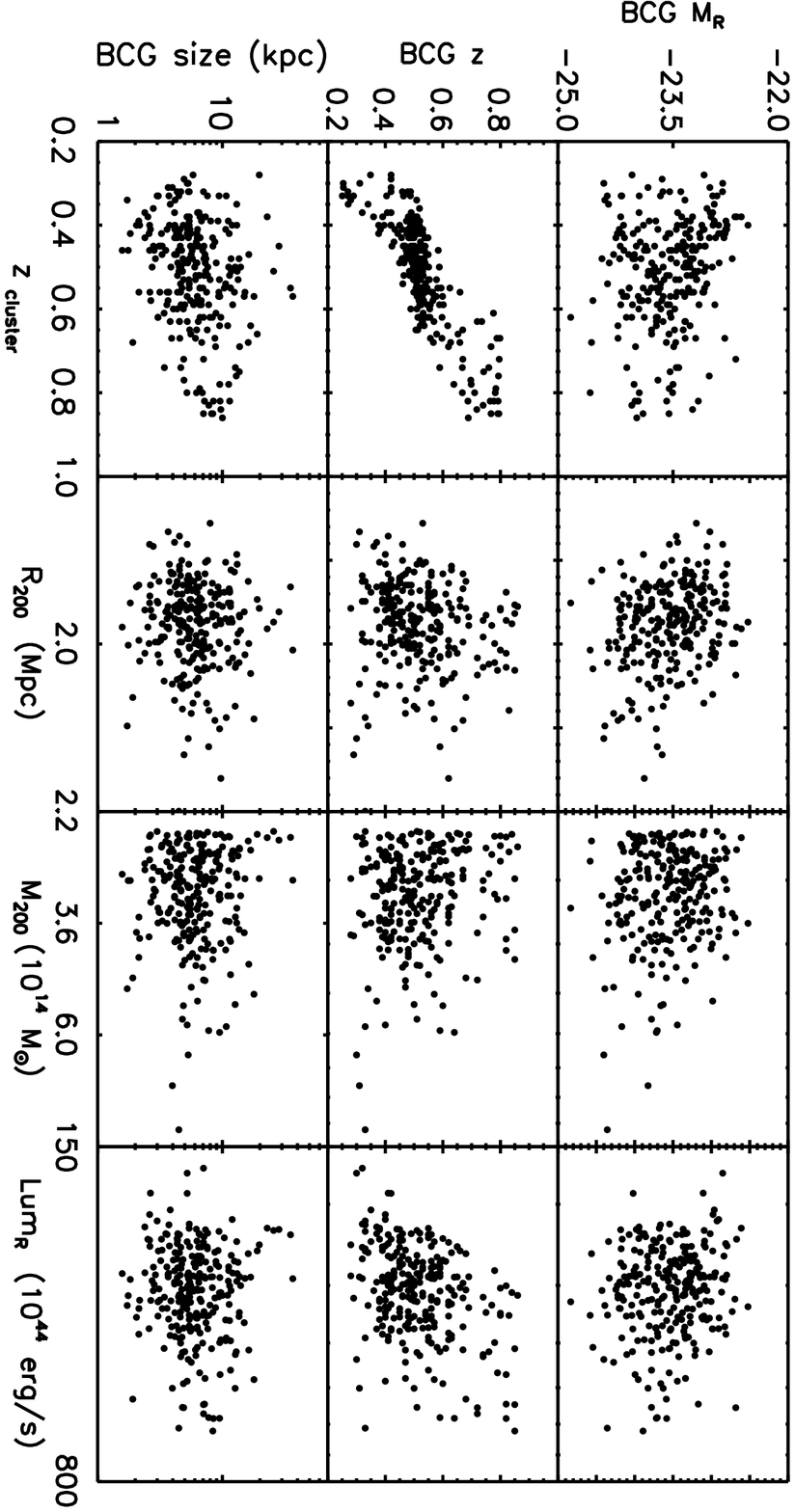} 
\caption{BCGs parameters against host cluster properties (see landscape).}
\label{fig:BCGhost}
\end{figure}

We now explore in more detail the significant correlation between the redshift of the cluster and the BCG absolute magnitude found.  In order to study this behavior for a fixed mass bin, we fit the evolution of the BCG luminosity with redshift for three masses bins: clusters with $2.4\times 10^{14} M_{\odot}<M_{200}<2.83\times 10^{14} M_{\odot}$, clusters with $2.83\times 10^{14} M_{\odot}<M_{200}<3.43\times 10^{14} M_{\odot}$ and clusters with mass larger than $3.43\times 10^{14} M_{\odot}$. These bins have been chosen in order to separate the sample into equal numbers. In Figure \ref{fig:zmag}, we can see the different fits to the BCG magnitude-redshift relation for different bins of mass.  The fits can be written as $M_R= A_0 +A_{1}z$, where $A_0$ and $A_1$ are listed in Table \ref{tab:Mrz}. 

\begin{table*}
      \caption{Results of the absolute magnitude versus redshift linear fit as a function of mass.}
      \[
         \begin{array}{ccc}
            \hline\noalign{\smallskip}
\multicolumn{1}{c}{\rm }&
\multicolumn{1}{c}{\rm A_0 }&
\multicolumn{1}{c}{\rm A_1 }\\
\hline\noalign{\smallskip}
\rm  2.4\times 10^{14} M_{\odot}  <M_{200}<2.83\times 10^{14}     &   -22.57 \pm 0.21 & -1.68  \pm 0.38   \\
\rm 2.83\times10^{14} M_{\odot} <M_{200}<3.43\times 10^{14} M_{\odot}        & -23.12 \pm 0.22 &  -1.18 \pm 0.40   \\
\rm 3.43\times 10^{14} M_{\odot}>M_{200}   &    -23.15 \pm 0.25 & -1.00 \pm 0.49  \\
         \end{array}
      \]
\label{tab:Mrz}
   \end{table*}

For comparison, \cite{wen12} reported a rate of evolution of the magnitude as a function of the redshift and richness based on their sample up to z$\sim$ 0.4.  \cite{wen12} provided fits for six different richness  ($R_L$) bins, finding a steeper slope for poorer clusters in concurrence with our results. Their two richest bins corresponds to masses between 2.42 and 3.89 $\times 10^{14}  M_{\odot}$ and $>$ 3.89 $\times 10^{14}  M_{\odot}$ respectively. Thus, for our two lowest mass bins, the \cite{wen12} results imply

\[ M_R = (-23.17 \pm 0.03) - (1.58 \pm 0.09)z, \]
and for our upper mass bin we use the fit
\[ M_R = (-23.44 \pm 0.04) - (1.35 \pm 0.17)z \]

The solid line in Fig. \ref{fig:zmag} refers to this fit with the richness estimated from each mass bin.  The slope in the magnitude-redshift relation obtained by  \cite{wen12} fully agrees with the slope that we find in a much broader range in redshift.  In addition, the evolution of the BCG magnitude with redshift  found in this work is also consistent with the one found by \cite{wen12} and extends linearly to higher redshift.

One possible effect that could affect these results could come from the fact that, according to the hierarchical scenario of structure formation, galaxy clusters are known to be less massive at higher redshift. In order to demonstrate that the chosen mass bins are narrow enough to be robust against this effect, we have color-coded the BCGs according to their host halo mass in Figure \ref{fig:zmag}. The two bottom panels (the two lower and narrower mass bins) do not show any trend of the mass with either redshift or magnitude and the correlation coefficients are consistent with zero. Therefore, we conclude that the observed redshift trend within each mass bin cannot be substantially affected by underlying trends of mass with redshift. The higher mass bin shows an absence of very massive clusters ($M>9\times10^{14}M_{\odot}$) at $z>0.5$ due to the real absence of such clusters according to the hierarchical scenario. In this case the significance of the correlation coefficient between the $M_{200}$ and the magnitude according to the Spearman test is 2.61, indicating that there could be a weak dependence. In this case, this effect might make the slope of the relation to look steeper, which does not seem to be the case.

We confirm the existence of the BCG magnitude evolution with redshift at fixed cluster mass extending the redshift range of previous results \citep{wen12}. Moreover, because we have shown that this cannot be a selection effect or a cluster mass evolution effect, this suggests that BCGs are passively evolving, at least within the redshift $0.25\le z \le 0.8$ and mass range  ($>2.4\times10^{14}M_{\odot}$) where our sample is complete. This result is in agreement with  other works supporting the hierarchical merging scenario  (i.e. \citealt{bildfell08,bernardi09}).

\section{Conclusions}

In this paper, we present the first optically selected cluster catalog in the DLS. We analyze the spatial, redshift and mass distribution of the detected clusters and compare the clusters with other optical, X-ray, WL and spectroscopic detections in the literature, finding generally good agreement for structures at z$\ge$ 0.25. When comparing with spectroscopic detections we find a good correlation for  the most massive structures and an increase in the distance and redshift offset of the matched structures for less massive detections. We also find, with high ($>3\sigma$) confidence,  that the optical and weak lensing maps are correlated, but with low correlation coefficients because the weak lensing maps are dominated by noise.  We also compare the masses estimated from optical richness and the \cite{dong08} mass-richness relation to masses estimated from WL or X-rays finding that the BCF provides a generally good estimator. In general, the DLS cluster set can be used as a tool to compare systematics between different methods. However, catalog users should be aware that very low redshift ($<$0.25) clusters are missing or biased upward in redshift because their member galaxies are saturated in the DLS imaging. 

Additionally, we inspected and visually classified the corresponding BCG for each cluster to investigate their properties. We restricted the DLS cluster sample to those with $M_{200}\ge2.4\times 10^{14} M_{\odot}$ within a redshift range of $0.25 \le z \le0.8$ to avoid biases. This is a wider range of mass and redshift than has been considered in previous work  \citep{ascaso11,wen12}. To understand the processes that lead to the formation of the BCG, we examine the relation between BCG redshift and absolute magnitude for different cluster masses. We confirm  the evolution of the BCG luminosity with redshift at fixed cluster mass observed by \cite{wen12} and extend it to higher redshift. In addition, we fully agree with the slope given by \cite{wen12}, suggesting that BCGs in clusters with masses $\ge 2.4\times 10^{14} M_{\odot}$ have passively evolved since redshift at least $\sim$ 0.8 in agreement with the hierarchical scenario \citep{delucia07}. On the other hand, other observational works \citep{vonderlinden07,lin10} have claimed a difference between BCGs in low-mass clusters and groups and high-mass clusters, with the former being more similar to normal ellipticals. Less massive clusters had formed later in time, not having had time to establish a cool core.

Consequently, even though we have shown evidence in this study that BCGs in massive clusters ($M_{200}\ge2.4\times 10^{14} M_{\odot}$)  evolve passively after redshift $\sim$ 0.8, we need larger and complete samples of BCGs down to lower cluster masses to shed light on the primary mechanisms driving evolution in low mass clusters and groups. Next generation optical surveys expected to have large deep areas with very good photometric redshift resolution (e.g: J-PAS, \citealt{benitez09}; LSST, \citealt{lsst12}; DES, \citealt{des05}; Euclid, \citealt{laureijs11}, among others) will provide excellent datasets for such purposes.

\section*{Acknowledgments}

We thank the anonymous referee for his/her really useful and constructive comments which help us improved the paper significantly. We acknowledge Ami Choi, Sam Schmidt, Vera Margoniner, James Jee, Jim Bosch, Russell Ryan, Tony Tyson and Txitxo Ben\'itez for useful comments and suggestions. B.A. acknowledges the support from Junta de Andaluc\'ia, through the Excellence Project P08-TIC-3531 and the Spanish Ministry for Science and Innovation, through grants  AYA2010-22111-C03-01 and CSD2007-00060.

\onecolumn

\twocolumn

\appendix

\begin{figure}
\centering
\includegraphics[clip,angle=0,width=1.0\hsize]{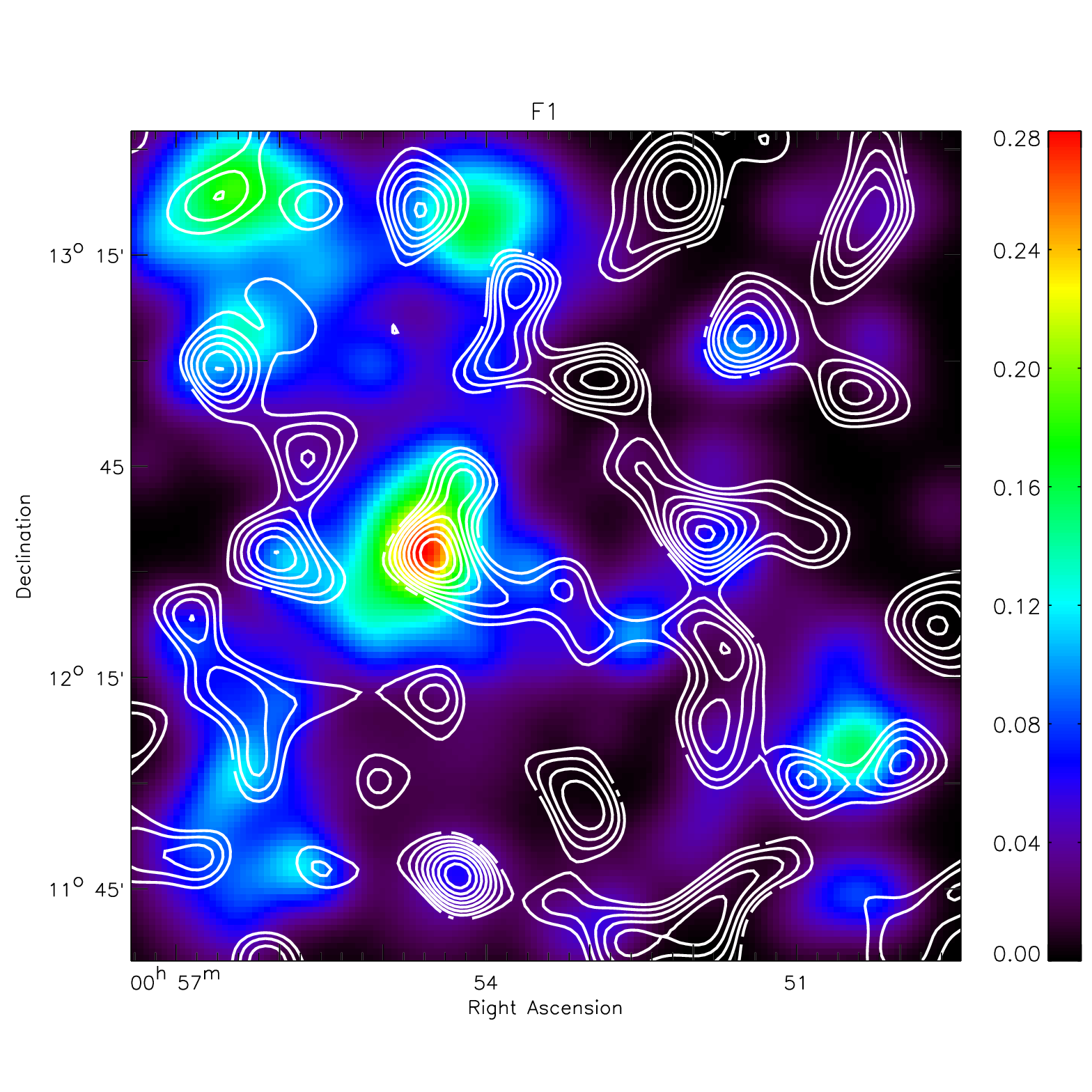} 
\caption{Synthetic weighted surface mass density map for the cluster detections in  DLS field F1 in linear scale (colour map) in arbitrary units. The white contours correspond to the smoothed WL signal-to-noise map with linear contour intervals spaced from 0 to 5. Regions without contours are underdense according to the lensing analysis.}
\label{fig:DLSWLcontF1}
\end{figure}

\begin{figure}
\centering
\includegraphics[clip,angle=0,width=1.0\hsize]{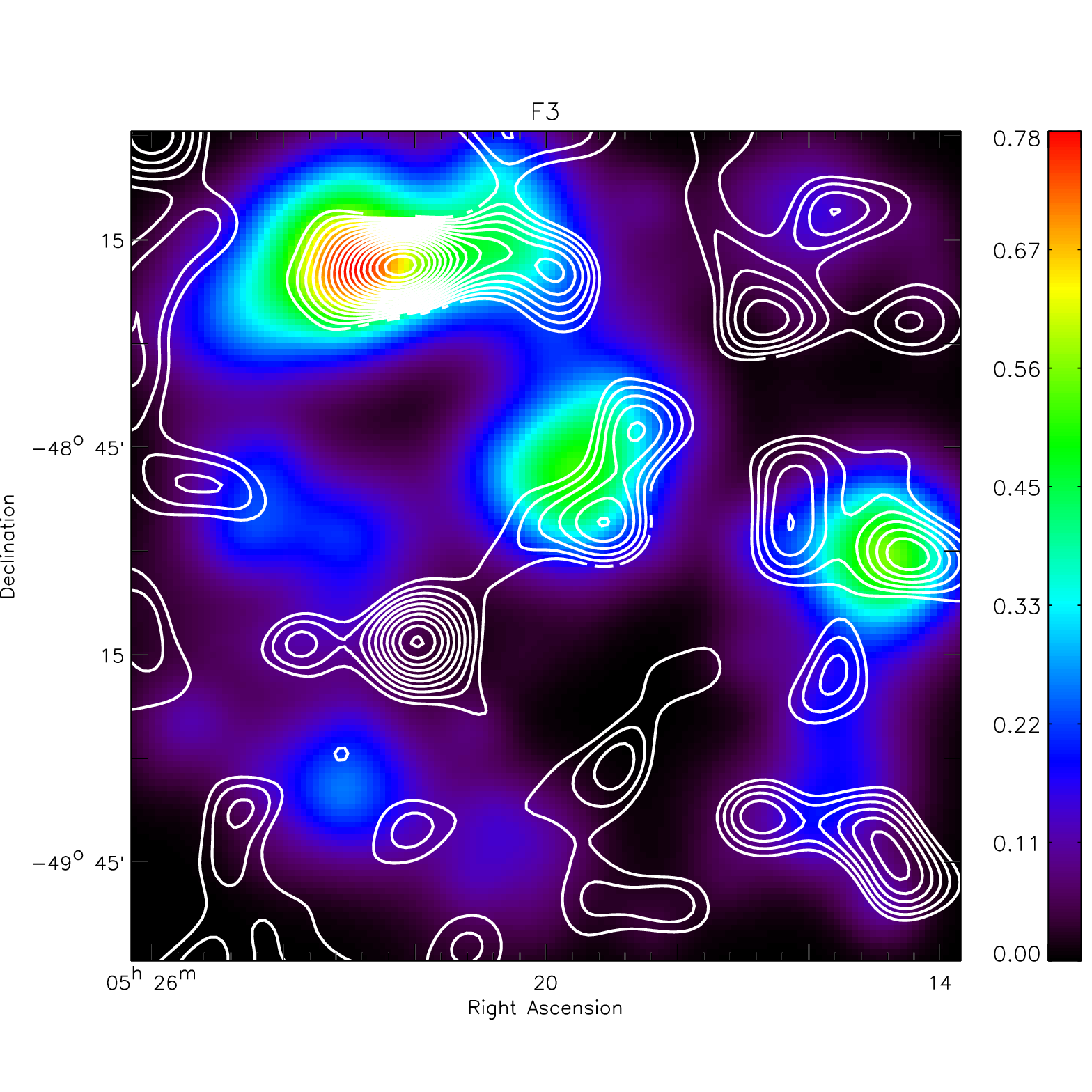} 
\caption{The same as in Fig \ref{fig:DLSWLcontF1} for the DLS field F3.}
\label{fig:DLSWLcontF3}
\end{figure}

\begin{figure}
\centering
\includegraphics[clip,angle=0,width=1.0\hsize]{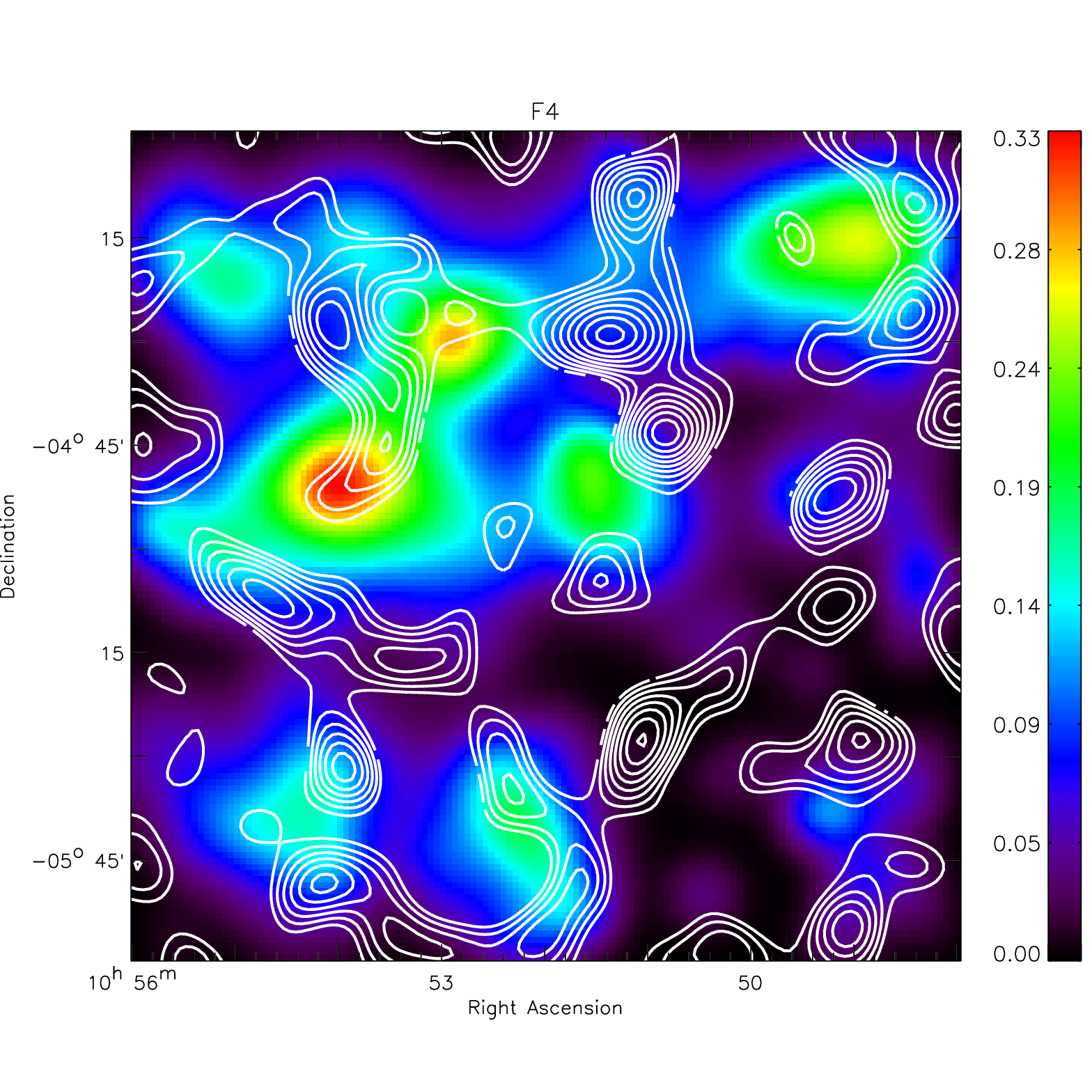} 
\caption{The same as in Fig \ref{fig:DLSWLcontF1} for the DLS field F4.}
\label{fig:DLSWLcontF4}
\end{figure}

\begin{figure}
\centering
\includegraphics[clip,angle=0,width=1.0\hsize]{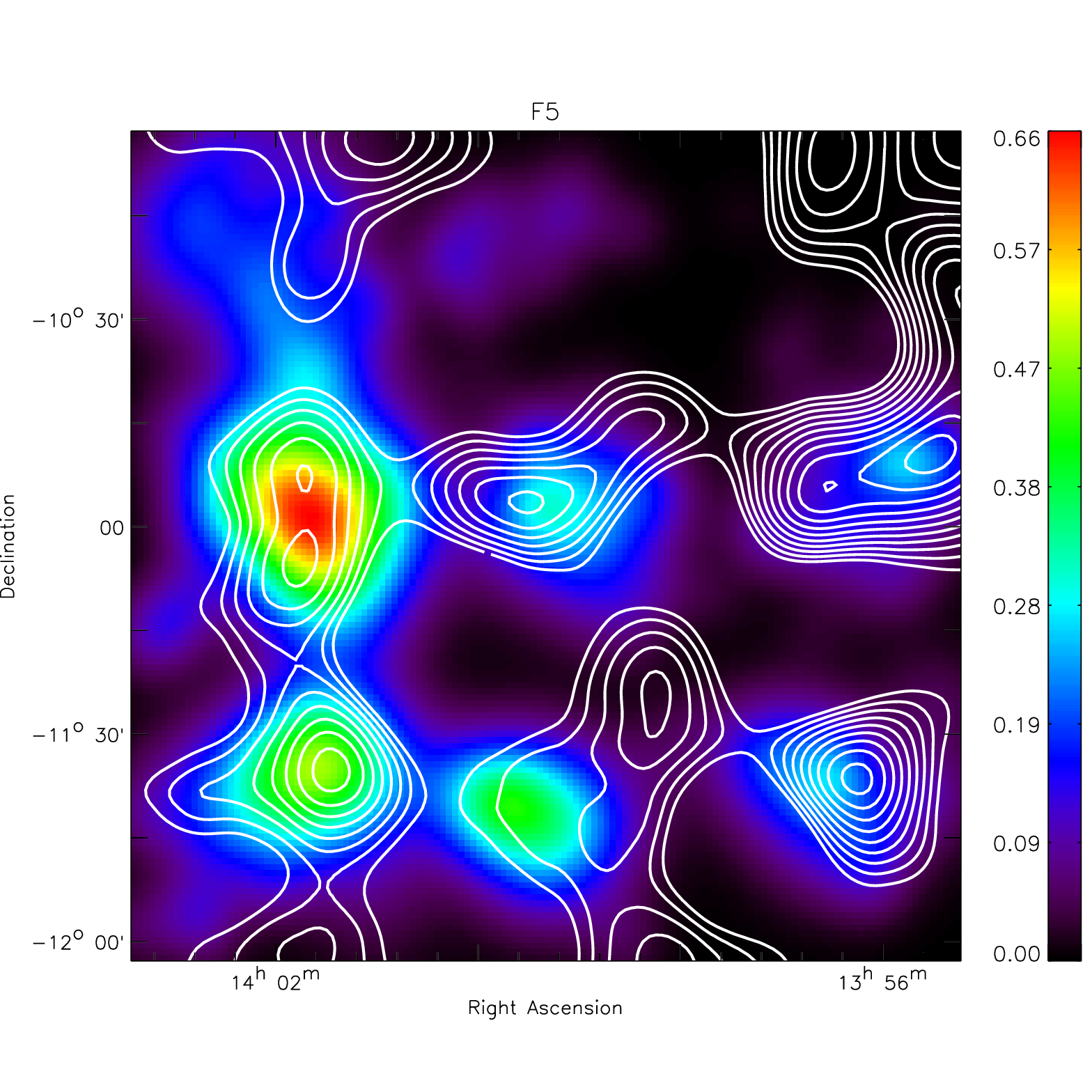} 
\caption{The same as in Fig \ref{fig:DLSWLcontF1} for the DLS field F5.}
\label{fig:DLSWLcontF5}
\end{figure}

\end{document}